\documentclass[12pt]{article}

\usepackage{amsmath}
\usepackage{amssymb}
\usepackage{amsthm}
\usepackage{relsize}
\usepackage{hyperref}
\usepackage{color}
\usepackage{mathrsfs}
\usepackage{graphicx}
\usepackage{bm}
\usepackage{pgfplots}
\usepackage{tikz}
\usepackage{booktabs}
\usepackage{verbatim}
\usepackage{hhline}
\usepackage{geometry}
\usepackage{multirow}
\usepackage{textcomp}
\usepackage{natbib}
\usepackage{authblk}
\usepackage{appendix}
\usepackage{tcai}

\newtheorem{prop}{Proposition}
\newtheorem*{remark*}{Remark}
\newcommand{\argmin}{\operatornamewithlimits{argmin}}

\newcommand{\ech}{\color{black}\rm}

\newcommand{\compn}[2]{\left(
\begin{array}{c}
#1\\
#2
\end{array}
\right)}

\newgeometry{left=3cm,bottom=2cm}

\title{Bayesian Nonparametric Ordination 
for   the  Analysis  of  Microbial Communities}
\author[1]{Boyu Ren}
\author[2]{Sergio Bacallado}
\author[3]{Stefano Favaro}
\author[4]{Susan Holmes}
\author[1]{Lorenzo Trippa}
\affil[1]{Harvard University, Cambridge, USA}
\affil[2]{University of Cambridge, Cambridge, UK}
\affil[3]{Universit\`a degli Studi di Torino and Collegio Carlo Alberto, Turin, Italy}
\affil[4]{Stanford University, Stanford, USA}

\begin{document}

\maketitle

\begin{abstract}
Human microbiome studies use sequencing technologies to measure the abundance of bacterial species or Operational Taxonomic Units (OTUs) in samples of biological material. Typically the data are organized in contingency tables with OTU counts across heterogeneous biological samples. In the microbial ecology community, ordination methods are frequently used to investigate latent factors or clusters that capture and describe variations of OTU counts across biological samples. It remains important to evaluate how uncertainty in estimates of each biological  sample's  microbial distribution propagates to ordination analyses, including visualization of clusters and projections of biological samples on low dimensional spaces. We propose a Bayesian analysis for dependent distributions to endow frequently used ordinations with estimates of uncertainty. A Bayesian nonparametric prior for dependent normalized random measures is constructed, which is marginally equivalent to the normalized generalized Gamma process, a well-known prior for nonparametric analyses. In our prior, the dependence and similarity between microbial distributions is represented by latent factors that concentrate in a low dimensional space. We use a shrinkage prior to tune the dimensionality of the latent factors. The resulting posterior samples of model parameters can be used to evaluate uncertainty in analyses routinely applied in microbiome studies. Specifically, by combining them with multivariate data analysis techniques we can visualize credible regions in ecological ordination plots. The characteristics of the proposed model are illustrated through a simulation study and applications in two microbiome datasets.
\end{abstract}

\noindent%
\vfill

\newpage
\section{Introduction}
\label{sec:intro}

Next generation sequencing (NGS) has transformed the study of microbial ecology.
Through the availability of cheap efficient amplification and sequencing,
marker genes such as 16S rRNA are used to provide inventories of
bacteria in many different environments. For instance soil and 
waste water microbiota have been inventoried \citep{greengenes} as well as the 
human body \citep{dethlefsen2007}. 
NGS also enables researchers to describe
the {\it metagenome} by computing counts of DNA reads and matching them to the 
genes present in various environments.

Over the last ten years,
numerous studies have shown 
the effects of  environmental and clinical factors on the 
bacterial communities of the human microbiome. These studies  enhance
our understanding of how the microbiome is involved in
obesity \citep{turnbaugh2009},
Crohn's disease \citep{quince2013Crohns}, or diabetes \citep{kostic2015dynamics}. Studies are currently 
underway to improve our understanding of the 
effects of
 antibiotics \citep{dethlefsen2011incomplete}, pregnancy \citep{PNAS2015}, and other perturbations
to the human microbiome.

Common microbial ecology pipelines either start by grouping the
16S rRNA sequences into known Operational Taxonomic Units (OTUs) or taxa
as done in 
\cite{qiime}, or denoising and grouping the reads into more refined strains
sometimes  referred to as oligotypes, phylotypes, or ribosomal variants (RSV)
\citep{DADA,eren2014oligotyping, dada2}.    We  will   call  all  types  of  
groupings  {OTU}s
to maintain consistency. In all cases the  data are analyzed in the form of
contingency tables of read counts per sample for the different OTUs , as exemplified in Table \ref{otu.table.example}.
Associated to these contingency tables are clinical and environmental 
covariates such as time, treatment, and patients' BMI, information  
collected on the same biological samples or environments. 
These are sometimes
   misnamed
``metadata''; this contiguous information is usually  fundamental  in the analyses.
The data  are often assembled in multi-type structures,
for instance  {\tt phyloseq } \citep{phyloseq} uses lists (S4 classes) to capture all the different aspects of the data at once.

Currently bioinformaticians and statisticians analyze the
preprocessed microbiome data using linear ordination methods such as 
Correspondence Analysis (CA), Canonical or Constrained Correspondence Analysis (CCA) , and 
Multidimensional Scaling (MDS) 
\citep{qiime,vegan,phyloseq}. 
Distance-based ordination methods use measures of between-sample or Beta diversity, such as the Unifrac distance \citep{unifrac}. 
These analyses can reveal clustering of biological samples or taxa, or
meaningful ecological or clinical gradients 
   in the community structure of the bacteria.
   Clustering, when it occurs indicates  a latent variable which is discrete,
 whereas gradients correspond to latent continuous variables.
Following these exploratory stages, confirmatory analyses can include
differential abundance testing \citep{wastenotwantnot},
 two-sample tests for Beta diversity scores \citep{beta-test}, ANOVA permutation tests in CCA \citep{vegan}, or tests based on  generalized linear  models that include  adjustment  for multiple confounders \citep{metagenomeseq}.

The interaction between these tasks can be problematic. In particular, the uncertainty in the  estimation of OTUs' prevalence
 is often not propagated to subsequent steps \citep{Peiffer}. Moreover,
   unequal sequencing depths 
  generate   variations of  the number of OTUs  with zero counts across biological samples.
  Finally, the hypotheses tested in the inferential step are often formulated after significant exploration of the data and are sensitive to earlier choices in data preprocessing. 

These issues motivate a Bayesian approach that  enables 
us to integrate the steps of the analytical pipeline.
 \cite{holmes2012dirichlet,la2012hypothesis,ding2014dynamics} have suggested the use of a simple Dirichlet-Multinomial
 model  for these data; however, in those analyses the multinomial probabilities for each biological sample are independent in the prior and posterior, which fails to capture underlying relationships between biological
 samples. The simple  Dirichlet-Multinomial model is also not able to account for  strong positive correlations (high co-occurrences
\citep{faust2012microbial}) or negative correlations (checker board effect \citep{koenig2011}) that can
exist between different species \citep{gorvitovskaia2016}.

We propose a Bayesian procedure, which  jointly  models the read counts from different OTUs  and  sample-specific latent multinomial distributions,  allowing
for   correlations between OTUs. 
 The prior assigned to these multinomial probabilities is highly flexible, such that the analysis learns the dependence structure from the data, rather than constraining it {\it a priori}. 
The method can deal with uncertainty coherently, provides model-based visualizations of the data, and is extensible to describe the effects of observed clinical and environmental covariates.

Bayesian analysis with Dirichlet priors is a convenient starting point for microbiome data, since the OTU distributions are inherently discrete. Moreover, Bayesian nonparametric priors for discrete distributions,  suitable for an unbounded number of OTUs, have been the topic of intense research in recent years. General classes of priors such as normalized random measures have been developed, and their properties in relation to classical estimators of species diversity are well-understood \citep{ferguson,hjort2010bayesian}. The problem of modeling dependent distributions has also been extensively studied  since the proposal of the Dependent Dirichlet Process \citep{maceachern2000dependent} by  \cite{muller2004}, \cite{rodriguez}, and \cite{griffin}). 


In this paper, we try to capture the variation in the composition of microbial communities as a result of  a group of unobserved samples' characteristics. With this goal  we introduce a model which expresses the dependence between OTUs  abundances  in different environments through vectors embedded in a low dimensional space.
 Our model has aspects in common with nonparametric priors for dependent distributions, including a generalized Dirichlet type marginal prior on each distribution, but is  also similar in spirit to the multivariate methods currently employed in the microbial ecology community. Namely, it allows us to visualize the relationship between biological samples 
 through low dimensional projections. 
 

The paper is organized as follows. Section 2 describes a prior for dependent microbial distributions, first constructing the marginal prior of a single discrete distribution through manipulation of a Gaussian process and then extending this to multiple correlated distributions. 
The extension is achieved through a set of continuous latent  factors, one for each biological sample, whose prior has been frequently used in Bayesian factor analyses. Section 3 derives an MCMC sampling algorithm for posterior inference and a fast algorithm to estimate biological samples' similarity. Section 4 discusses a method for visualizing the uncertainty in ordinations through conjoint analysis. Section 5 contains analyses of simulated data, which serve to demonstrate desirable properties of the method, followed by applications to real microbiome data in Section 6. Section 7 discusses potential improvement and concludes. The  code  for  implementing  the analyses  discussed in this  article is included in
the Supplementary Materials.


\section{Probability Model}
In Table \ref{otu.table.example}, we illustrate an example of a typical OTU table with 10 biological samples, where half  are healthy subjects, and half are Inflammatory Bowel disease (IBD) patients.
 This contingency table is a subset of the data in \cite{morgan2012dysfunction} and records the observed frequencies of five most abundant genus level OTUs in all biological samples based on 16S rRNA sequencing results.
 Let $Z_i$ be the $i$th observed OTU (e.g. $Z_1$ is Bacteroides) and $n_{i,j}$ be the observed frequency of OTU $Z_i$ in biological sample $j$. As an example, $n_{11}=1822$ is the observed frequency of Bacteroides in the biological sample Ctrl1. We will denote an OTU table as $(n_{i,j})_{i\leq I,j\leq J}$, where $I$ is the number of observed OTUs and $J$ the number of biological samples.

For the biological sample $j$, we will assume the vector $(n_{1,j},\ldots,n_{I,j})$ follows a multinomial distribution, noting that our analysis extends easily to the case in which the total count $\sum_{i=1}^I n_{ij}$ is a Poisson random variable.
The unobserved multinomial probabilities of OTUs present in biological sample $j$ determine the distribution of the frequencies $n_{i,j}$. 
These probabilities form a discrete probability measure, which we call a microbial distribution, on the space $\Zsc$ 
of all OTUs.

We denote this discrete measure as $P^j$ and $P^j(\{Z_i\})$ gives the probability of sampling $Z_i$ from biological sample $j$. If we consider all $J$ biological samples, we expect there will be variation in the respective $P^j$'s. This variation usually can be explained by specific characteristics of the biological sample. For instance, in Table \ref{otu.table.example}, we can see the empirical multinomial probability of Enterococcus is higher in healthy controls than in IBD patients on average. This variation has  been discovered in prior publication \citep{morgan2012dysfunction} and is attributed to the IBD status. Microbiome studies aim to elucidate the characteristics that explain these types of variations.
\begin{table}
\footnotesize
\centering
\caption{\footnotesize{An example of OTU table derived from data published in \cite{morgan2012dysfunction}.}}
\label{otu.table.example}
\begin{tabular}{@{}ccccccccccc@{}}
\toprule
OTU             &Ctrl1&Ctrl2&Ctrl3&Ctrl4&Ctrl5&IBD1&IBD2&IBD3&IBD4&IBD5\\ \midrule
Bacteroides     & 1822      & 913       & 147       & 2988            & 4616      & 172   & 3516  & 657   & 550   & 1423    \\
Bifidobacterium& 0         & 162       & 0         & 0              & 84        & 0     & 85    & 1927  & 0     & 286        \\
Collinsella     & 1359      & 0         & 0         & 206              & 0         & 327   & 0     & 0     & 160   & 122       \\
Enterococcus    & 621       & 0         & 0         & 3                  & 40        & 0     & 0     & 0     & 0     & 0         \\
Streptococcus   & 75        & 139       & 2161      & 110              & 97        & 1820  & 85    & 58    & 5     & 294      \\
\bottomrule
\end{tabular}

\end{table}

Our method focuses on modeling the  distributions $P^j$'s and the variations among them. For biological samples labelled 
in $\mathcal J= \{1,\dots,J\}$,
we assume they have the same infinite set of OTUs $Z_1,Z_2,\ldots\in\Zsc.$ We let the number of OTUs present in a biological sample be infinity to make our model nonparametric in consideration of the fact that there might be an unknown number of OTUs that are not observed in the experiment. We specify the probability mass assigned to a group of OTUs $A\subset \Zsc$ as
\begin{equation}
\begin{aligned}
P^j(A) &= M^j(A)/M^j(\Zsc),\\
M^j(A) &= \sum_{i=1}^{\infty}\mathbb I(Z_i\in A)\sigma_i\langle\bX_i,\bY^j\rangle^{+2},
\label{rmeasure}
\end{aligned}
\end{equation}
where $\sigma_i\in (0,1)$, $\bX_i,\bY^j\in \mathbb R^m$,  $\mathbb I(\cdot)$ is the indicator function, and $x^+ = x\times \mathbb I(x>0)$. In addition, $\langle\cdot,\cdot\rangle$ is the standard inner product in $\mathbb R^m$.

In this model specification, $\sigma_i$ is related to the average abundance of OTU $i$ across all biological samples. When $\sigma_i$ is large, the average probability mass assigned to OTU $Z_i$ will also be large. We refer to $\bX_i$ and $\bY^j$ as OTU vector and biological sample vectors respectively. The variation of the $P^j$'s is determined by the vectors $\bY^j$, which can be treated as latent characteristics of the biological samples that associate with microbial composition; for example, an unobserved feature of the subject's diet, such as vegetarianism, could affect the abundance of certain OTUs. We assume there are $m$ such characteristics, and the $l$th component in $\bY^j$ is the measurement of the $l$th latent characteristic in biological sample $j$. The vector $\bX_i$ denotes the effects of each of the $m$ latent characteristics on the abundance of the OTU $Z_i$. Therefore $\bX_i$ has $m$ entries.

In  subsection \ref{DPcon} we consider a single microbial distribution $P^j$ with fixed parameter $\bY^j$ and define a prior on $\bsigma=(\sigma_1,\sigma_2,\dots)$ and $(\bX_i)_{i\geq 1}$ which makes $P^j$ a Dirichlet process \citep{ferguson}. The degree of similarity between the discrete distributions  $\{P^j; j\in\mathcal J\}$ is summarized by the Gram matrix $(\phi(j,j') = \langle \bY^j,\bY^{j'}\rangle; j,j'\in \mathcal{J})$. Subsection \ref{second.subsec} discusses the  interpretation of  this matrix. Subsection \ref{third.subsec} proposes a prior for the parameters $\{\bY^j, j\in\mathcal J\}$ which has been  previously  used in  Bayesian factor analysis, and which has the effect of shrinking the dimensionality  of the Gram matrix $(\phi(j,j'))$ and is used to infer the number of latent characteristics $m$. The parameters $\{\bY^j, j\in\mathcal J\}$ or $(\phi(j,j'))$ can be used to visualize and understand variations  of  microbial distributions across biological samples.

\subsection{Construction of a Dirichlet Process}\label{DPcon}
The prior on $\bsigma=(\sigma_1,\sigma_2,\ldots)$ is the distribution of ordered points ($\sigma_i>\sigma_{i+1}$) in a Poisson process on $(0,1)$ with intensity
\begin{equation}
\nu(\sigma) = \alpha \sigma^{-1}(1-\sigma)^{-1/2},
\label{int.poisson}
\end{equation}
where  $\alpha>0$ is a concentration parameter.
Denote the index of component of $\bY^j$ and $\bX_i$ as $l$. 
Fix $j$, and let $\bY^j=(Y_{l,j},l\leq m)$ be a fixed vector in $\mathbb R^m$ such that $\langle\bY^j,\bY^j\rangle=1$. We let $\bX_i = (X_{l,i},l\leq m)$ be a random vector for $i=1,2,\ldots$ and $X_{l,i}$ be independent and $N(0,1)$ {\it a priori} for $l=1,2,\ldots,m$ and $i=1,2,\ldots$ Finally, let $G$ be a nonatomic probability measure on the measurable space $(\mathcal{Z},\mathcal F)$, where $\mathcal F$ is the sigma-algebra on $\Zsc$, and $Z_1,Z_2,\ldots$ is a sequence of independent random variables  with  distribution $G$. We claim that the probability distribution $P^j$ defined in Equation (\ref{rmeasure}) is a Dirichlet Process with base measure $G$. 

We note that the point process $\bsigma$ defines an infinite sequence of positive numbers,
the products  $\langle\bX_i,\bY^j\rangle$\ech, $i=1,2,\ldots$, are independent  Gaussian $N(0,1)$ variables,
and that the intensity $\nu$  satisfies the inequality  $\int_0^1 \sigma d\nu <\infty$.
 These facts directly imply that with probability 1,  $0<M^j(A)<\infty$  when $G(A)>0$.
      It also follows that for any sequence of disjoint sets $A_1,A_2,\ldots\in \Fsc$ the corresponding random variables  $M^j(A_i)$'s  are  independent. In different words,  $M^j$  is a completely random measure \citep{kingman}. The marginal L\'evy intensity  can be  factorized  as
$
\mu_M(ds) \times G(dz),
$
where
\begin{align*} \displaystyle
\mu_{M}(ds) &\propto \int_{0}^1\nu(\sigma)\left(\frac{1}{\sigma}\right)^{1/2}s^{-1/2}\exp \left(-\frac{s}{2\sigma}\right)d\sigma \;ds \\
&\propto \frac{\exp(-s/2)}{s}ds,\quad  \text{for } s\in (0,\infty).
\end{align*}
The above expression shows that  $M^j$  is a Gamma process.
We recall that   the  L\'evy  intensity  of  a  Gamma  process  is  proportional to the map $s\mapsto \exp(- c \times s) \times s^{-1}$\ech, where $c$  is  a positive  scale parameter.
In \citet{ferguson}  it  is shown   that a Dirichlet process can be  defined by normalizing a Gamma process. 
It  directly  follows that  $P^j$ is a Dirichlet Process with base measure $G$. 

\begin{remark*}
Our construction can be extended to a
  wider class of normalized random measures \citep{james2002poisson,regazzini}
   by  changing the intensity $\nu$  that  defines  the Poisson process $\bsigma$.
    If  we  set $$\nu(\sigma)= \alpha\sigma^{-1-\beta}(1-\sigma)^{-1/2+\beta}, $$
     $\beta \in [0,1)$, in  our
    definition  of  $M^j$ , then
 the     L\'evy   intensity  of  the  random  measure in (\ref{rmeasure})  becomes proportional  to
$$
s^{-1-\beta}\exp(-s/2).
$$
In this  case  the  L\'evy   intensity  indicates  that  $M^j$   is a  generalized Gamma process \citep{brix}.
 We  recall  that  by normalizing this class    one  obtains normalized generalized Gamma processes \citep{lijoi2007}, which include
 the  Dirichlet process and  the normalized Inverse Gaussian process \citep{InverseGaussian} as special cases.
\end{remark*}

A few  comments  capture  the  relation  between our  definition of   $P^j(A)$  in  (\ref{rmeasure})  and  alternative  definitions  of  the  Dirichlet  Process.  
 If  we  normalize  $h$ independent 
   $\text{Gamma}(\alpha/h,1/2) $  variables, 
   we obtain  a  vector with   $\text{Dirichlet}(\alpha/h,\ldots,\alpha/h)$ distribution.
      To  interpret  our  construction  we can note  that,
      when  $\alpha/h<1/2$,   each  of  the  $\text{Gamma}(\alpha/h,1/2) $  components
      can be  obtained  by  multiplying    a  $\text{Beta}(\alpha/h,1/2-\alpha/h)$ variable  and  an  independent
       $\text{Gamma}(1/2 , 1/2) $.
         The  distribution  of  the    $\langle\bX_i,\bY^j\rangle^{+2} $ 
      variables in (\ref{rmeasure})  is  in fact a mixture  with  a $\text{Gamma}(1/2 , 1/2) $     component  and   a  point  mass  at  zero.
       Finally if we  let  $h$  increase  to  $\infty$,   the  law  of the  ordered
      $\text{Beta}(\alpha/h,1/2-\alpha/h)$  converges weakly  to the law of ordered points of  a  Poisson point process  on $(0,1)$  with  intensity  $\nu$ (see Supplementary Document S1).
 
\subsection{Dependent Dirichlet Processes}
\label{second.subsec}
We use the   representation for Dirichlet processes  from  Equation (\ref{rmeasure})  to  define  a  family  of  dependent  Dirichlet processes labelled by a general index set $\Jsc$. The dependency structure of this family is related to $(\phi(j,j')=\langle\bY^j,\bY^{j'}\rangle)_{j,j'\in\Jsc}$. Geometrically $\phi(j,j')$ is the cosine of the angle between $\bY^{j}$ and $\bY^{j'}$.
%
The dependent Dirichlet processes is  defined  by  setting
\begin{equation}
P^{j}(A) = \frac{\sum_i \mathbb I(Z_i\in A)\times \sigma_i\langle\bX_i,\bY^{j}\rangle^{+2}}{\sum_i \sigma_i\langle\bX_i,\bY^{j}\rangle^{+2}},  \qquad  \forall j\in \mathcal{J},
\label{DP_mul}
\end{equation}
for  every  $A\in \mathcal{F}$. Here  the sequence  $(Z_1,Z_2,\ldots)$ and the array $(\bX_1,\bX_2,\ldots)$, as  in Section \ref{DPcon},  contain    independent  and  identically  distributed  random  variables, while  $\bsigma$  is  our Poisson
process on the unit interval defined in (\ref{int.poisson}). 
 We   will  use  the notation $Q_{i,j}=\langle\bX_i,\bY^{j}\rangle$.
This construction  has an  interpretable dependency  structure  between  the  $P^{j}$'s   that  we state in the  next  proposition.
\begin{prop}
\label{DP_mul_p}
There  exists  a real function   $\eta: [0,1]\rightarrow [0,1]$ such  that  the
correlation between $P^{j}(A)$ and $P^{j'}(A)$ is  equal  to $\eta \left( \phi(j,j') \right )$  for  every  $A$  that  satisfies  $G(A)>0$. In  different  words,   the correlation between $P^{j}(A)$
 and $P^{j'}(A)$ does  not  depend  on  the  specific  measurable  set $A$,
it is  a function of the angle  defined by  $\bY^{j}$  and $\bY^{j'}$.
\end{prop}

\begin{figure}[htbp]
\centering
\includegraphics[scale=0.33]{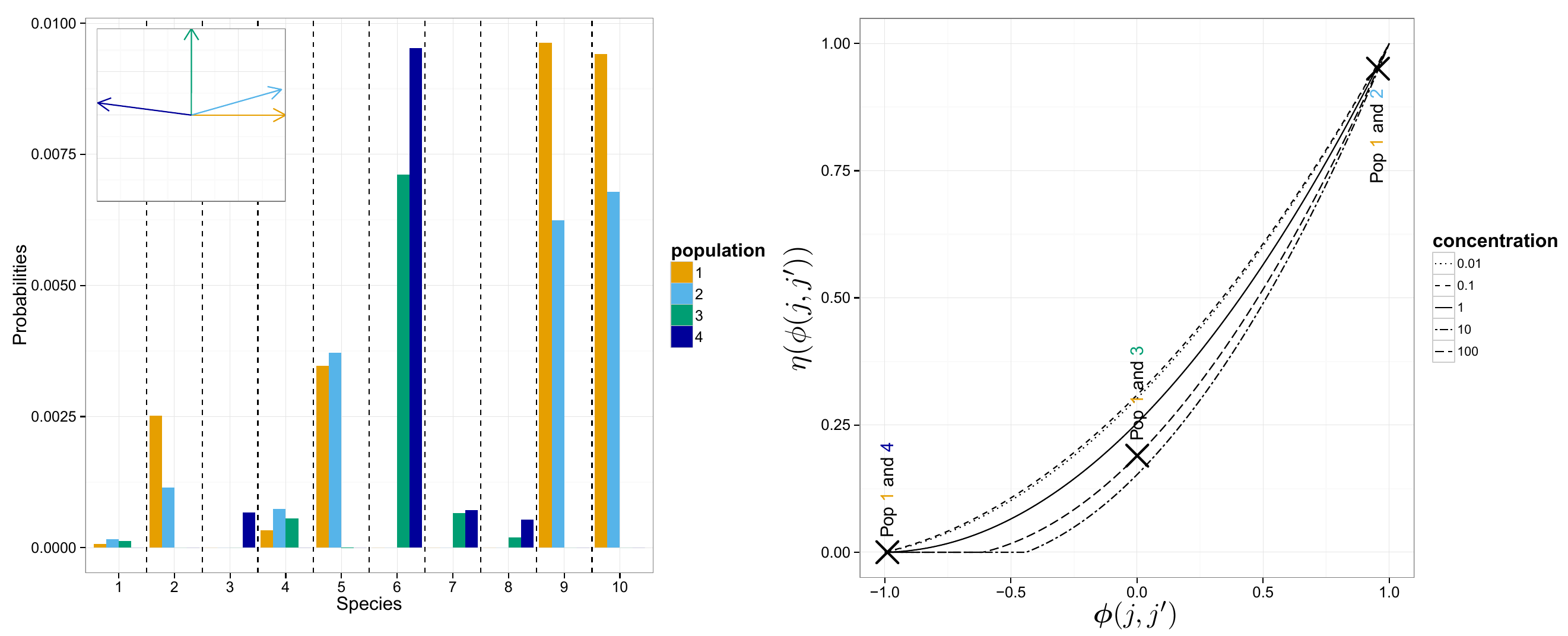}
\caption{\footnotesize{(\textbf{Left}) Realization of 4 microbial distributions from  our dependent Dirichlet processes.
 We  illustrate 10  representative  OTUs    and set $\alpha=100$.
  The miniature figure at the top-left corner shows the relative positions of the four biological sample vectors $\bY^j$. 
  The OTUs are those  associated  to the  10 largest $\sigma$'s.  
  As  suggested  by  this  panel,   the larger the angle between two $\mathbf{Y}^j$'s, the more  the  corresponding  
    random  distributions  tend  to  concentrate  on  distinct  sets. 
  (\textbf{Right}) Correlation of two random probability  measures 
  when the  cosine $\phi(j,j')$ between $\bY^{j}$ and $\bY^{j'}$  varies from $-1$ to $1$. 
    We consider    five different values of the concentration parameter $\alpha$.
  In  the  right  panel we also   mark  with crosses  the  correlations   between   $P^j(A)$ and $P^{j'}(A)$    for   pairs of  biological samples $j,j'$  considered   in the  left panel.}}
\label{DDP.fig}
\end{figure}

The proof is in the Supplementary Document S2. The first panel of  Figure  \ref{DDP.fig} shows  a  simulation of $P^{j}$'s. In this  figure  $\mathcal{J}=\{1,2,3,4\}$. When $\phi(j,j')$, the cosine of the angle between
two  vectors $\bY^j$ and $\bY^{j'}$, corresponding  to  distinct  biological  samples $j$ and $j'$, decreases  to $-1$ the   random measures tend to concentrate  on two   disjoint sets. The second panel  shows  the function  $\eta$  that  maps the  $\phi(j,j')$'s  into  the  correlations  $\corr(P^{j}(A), P^{j'}(A))=\eta(\phi(j,j'))$. As  expected the correlation  increases  with  $\phi(j,j')$.

We want to point out that the construction in (\ref{DP_mul}) extends easily to the setting where we are given any positive semi-definite kernel $\phi:\mathcal{J}\times\mathcal{J}\to(-1,1)$ capturing the similarity between biological samples labelled by $\Jsc$. Mercer's theorem \citep{mercer1909functions} guarantees the kernel is represented by the inner product in an $\Lsc^2$ space, whose elements are infinite-dimensional analogues of the vectors $\bY^j$. The analysis presented in this section is unchanged in this general setting.

The  next  proposition  provides  mild  conditions that guarantee  a  large  support for   the    dependent  Dirichlet  processes  that we  defined.
 %
\begin{prop}

Consider a collection of  probability measures $(F_j,j=1,\ldots,J)$  on $\Zsc$ and a positive  definite  kernel $\phi$. 
Assume that $\Jsc=\{1,\ldots,J\}$ and  the  support  of $G$ coincides with  $\Zsc$. 
The   prior distribution   in  (\ref{DP_mul}) assigns strictly positive probability  to  
the neighborhood 
$
\{(F'_1,\ldots,F'_J): |\int f_idF'_j -\int f_idF_j|<\epsilon, i=1,\ldots,L,~j=1,\ldots,J\}
$, where 
  $\epsilon>0$ and  $f_i$, $i=1,\ldots,L$,  are  bounded  continuous  functions. 
\end{prop}

In what  follows we will  replace  the constraint  $\langle\bY^j,\bY^j\rangle=1$   with  the requirement  $\langle\bY^j,\bY^j\rangle < \infty$. The two constraints  are equivalent  for  our  purpose, because  we normalize $M^{j}(\cdot) = \sum_i \mathbb I(Z_i\in \cdot)\times \sigma_i\langle\bX_i,\bY^{j}\rangle^{+2}$,  and     $\langle\bY^j,\bY^j\rangle$  can be  viewed  as a scale  parameter.

\subsection{Prior on biological sample parameters}
\label{third.subsec}

This subsection deals with the task of estimating  the parameters $\bY^j, j\in\mathcal J= \{1,\dots,J\}$, that capture most of the variability observed when comparing $J$ biological samples with different OTU counts. We define a joint prior on these factors which makes them concentrate on a low dimensional space; equivalently, the prior tends to shrinks the nuclear norm of the Gram matrix $(\phi(j_1,j_2))_{j_1,j_2\in\mathcal J}$. The problem of estimating low dimensional factor loadings or a low-rank covariance matrix is common in Bayesian factor analysis, and the prior defined below has been used in this area of research.

The parameters $\bY^j$ can be interpreted as key characteristics of the biological samples that affect the relative abundance of OTUs. As in factor analysis, it is difficult to interpret these parameters unambiguously \citep{press,rowe2002}; however, the angles between their directions have a clear interpretation. As observed in Figure \ref{DDP.fig}, if the kernel $\phi(j_1,j_2)\approx\sqrt{ \phi(j_1,j_1)\phi(j_2,j_2)}$, the two microbial distributions $P^{j_1}$ and $P^{j_2}$ will be very similar. If $\phi(j_1,j_2)\approx 0$, then there will be little correlation between OTUs' abundances in the two samples. If $\phi(j_1,j_2)\approx -\sqrt{ \phi(j_1,j_1)\phi(j_2,j_2)}$, then the two microbial distributions are concentrated on disjoint sets. This interpretation suggests Principal component analysis (PCA) of the Gram matrix $(\phi(j_1,j_2))_{j_1,j_2\in\mathcal J}$ as a useful exploratory data analysis technique. 

It is common in factor analysis to restrict the dimensionality of factor loadings. In our model, this is accomplished by assuming $\bY^j$ to be in $\mathbb R^m$ and adding an error term $\epsilon$ in the definition of $Q_{i,j}$, the OTU-specific latent weights,
\begin{equation}
Q_{i,j} = \langle \bX_i,\bY^{j}\rangle + \epsilon_{i,j},
\label{factor.model}
\end{equation}
where the $\epsilon_{i,j}$ are independent standard normal variables. Recall that each sample-specific random distribution $P^j$ is obtained by normalizing the random variables $\sigma_i (Q_{i,j} ^+ ) ^2$. If we denote the covariance matrix of $(Q_{i,1},\ldots,Q_{i,J})$ as $\bSigma$, this factor model specification indicates $\bSigma = \bY^\intercal\bY + \bI$ conditioning on $\bY$, where $\bI$ is the identity matrix and $\bY = (\bY^1,\ldots,\bY^J)$. As a result, the correlation matrix $\bS$ induced by $\bSigma$ only depends on $\bY$.

In most applications  the  dimensionality  $m$  is  unknown. Several approaches to  estimate  $m$  have been  proposed  \citep{lopes2004,lee2002,lucas2006,carvalho2008,ando2009}.
However, most of them involve either calculation of Bayes Factors or complex MCMC algorithms.
Instead  we   use a normal shrinkage prior proposed by \cite{dunsonfactor}.
This prior includes an  infinite   sequence  of factors ($m=\infty$),  but   the   variability   captured   by  this sequence of  latent  factors   rapidly  decreases  to  zero.
 A  key  advantage of the model  is  that  it  does not  require the user to choose the number of factors.
The  prior  is  designed  to  replace  direct selection  of  $m$  with  the  shrinkage
 toward  zero  of the
  unnecessary  latent  factors.
 In addition, this prior  is  nearly conjugate, which simplifies  computations. The prior  is defined as follows,
\begin{equation}
\begin{aligned}
&\gamma_l\sim \text{Gamma}(a_l,1), \qquad
\gamma'_{l,j}\sim \text{Gamma}(v/2,v/2),\\
&Y_{l,j}|\bgamma\sim N\left(0,(\gamma_{l,j}')^{-1}\prod_{k\leq l}\gamma_k^{-1}\right),\;~l\geq 1,~j\in \Jsc,
\end{aligned}
\label{dunson_prior}
\end{equation}
where  the  random  variables  $\bgamma=(\gamma_l,  \gamma'_{l,j};  \; l,j\ge 1)$  are independent and,  conditionally  on these  variables,  the $Y_{l,j}$'s
 are independent.

\begin{figure}[htbp]
\centering
\includegraphics[scale=0.7]{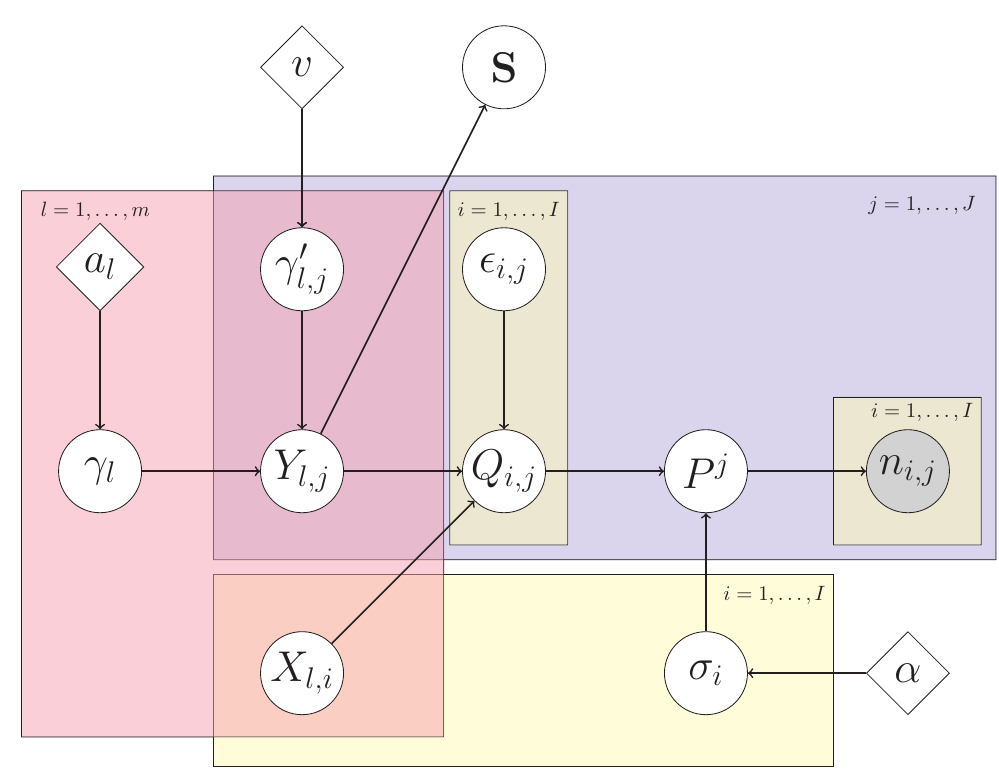}
\caption{\footnotesize{Plate diagram. We include the factor model for the latent variables $Q_{i,j}$ as well as the matrix $\bS$. Nodes encompassed by a rectangle are defined over the range of indices indicated at the corner of the rectangle, and the connections shown within the rectangle are between nodes with the same index. We use $j$ to index biological samples, $i$ to index microbial species and $l$ to index the components of latent factors.}}
\label{plate.diagram}
\end{figure}

When $a_l>1$,  the  shrinkage  strength  {\it a  priori} 
increases  with  the  index  $l$,  and  therefore the variability captured  by  each  latent factor  tends  to  decrease  with $l$.
  We refer  to  \cite{dunsonfactor}  for  a detailed  analysis of  the  prior in  (\ref{dunson_prior}). 
In practice,  the assumption of infinitely many factors  is  replaced  for    data analysis  and posterior  computations   by a finite  and  sufficiently  large number  $m$ of factors. The  choice  of  $m$ is  based  on computational  considerations. It  is  desirable  that  posterior  variability of  the   last  components ($l\sim m$)  of  the  factor  model in (\ref{factor.model}) is negligible.
This  prior model  is   conditionally  conjugate when paired with the dependent Dirichlet processes prior in subsection \ref{second.subsec},  a relevant and convenient characteristic for posterior simulations. We summarize the full model with a plate diagram, shown in Figure \ref{plate.diagram}.

\section{Posterior  Analysis}
\label{sec.post}
Given an exchangeable sequence  $W_1,\ldots,W_n$   from   $P^j=M^j \times  M^j(\mathcal{Z})^{-1}$ as defined in subsection \ref{DPcon},  we can rewrite the likelihood function  using  variable  augmentation as in \cite{james2009},
\begin{equation}
\begin{aligned}&\prod_{i=1}^n  P^j(\{W_i\})=
\int_0^{\infty} 
 \frac{\exp[-M^j(\mathcal{Z})\;T] \times T^{n-1}}{\Gamma(n)} \prod_{i=1}^I  M^j(\{W_{i}^*\})^{n_i} dT.
\end{aligned}
\label{SSS_ll}
\end{equation}
Here $W_1^*, \ldots,W_I^*$  is  the  list  of  distinct  values  in  $(W_1,\ldots W_n)$  and  $n_1,\ldots,n_I$ are the occurrences in $(W_1,\ldots W_n)$, so that  $\sum_{i=1}^{I} n_i=n$.
  We  use  expression (\ref{SSS_ll})   to    specify   an  algorithm that  allows us to infer   microbial abundances  $P^{1}, \ldots, P^{J}$  in   $J$   biological samples.

We  proceed,  similarly to \cite{muliere1998} and \cite{ishwaran2001gibbs},  using  truncated  versions of  the   processes in subsection \ref{second.subsec}.
We replace   $\bsigma=\{\sigma_i,i\geq 1\}$
with  a   finite number $I$ of independent     $\text{Beta}(\epsilon_I,1/2-\epsilon_I)$
 points in  $(0,1)$.
     Supplementary Document S1 shows  that when $I$ diverges, and $\epsilon_I=\alpha/I$,  this   finite dimensional version  converges weakly to the  process  in (\ref{int.poisson}).
      Each  point  $\sigma_i$  is    paired with  a multivariate normal  $\bQ_i=(Q_{i,1}, \ldots, Q_{i,J})$  with mean zero  and  covariance  $\bSigma$.
    The  distribution  of   $M_{i,j} = \sigma_i (Q_{i,j}^+)^2$
     is  a mixture of a   point mass at  zero  and  a  Gamma  distribution.
           %
       In this section $\bQ$ and $\bsigma$ are finite dimensional,
and  the  normalized   vectors $P^{j}$, which  assign  random  probabilities   to    $I$ OTUs in $J$ biological samples,
         are   proportional  to $(M_{1,j},\ldots, M_{I,j}),$ $j=1,\ldots,J$.
          Note  that  $P^j$ conditional on $\mathbb I( Q_{1,j}>0),\ldots,\mathbb I( Q_{I,j}>0)  $  follows  a Dirichlet distribution with  parameters
         proportional to $\mathbb I( Q_{1,j}>0),\ldots,\mathbb I( Q_{I,j}>0)  $.


     The algorithm is  based on  iterative sampling  from  the full conditional  distributions.
 We first  provide a description  assuming that $\bSigma$ is known.
We    then  extend   the  description    to   allow   sampling   under  the shrinkage prior  in  Section $\ref{third.subsec}$ and
to infer $\bSigma$.

With $I$ OTUs and $J$ biological samples,  the typical  dataset
is   $\bn=(\bn_1,\ldots,\bn_J)$, where  $\bn_j=(n_{1,j},\ldots,n_{I,j})$ and $n_{i,j}$ is the absolute frequency of  the $i$th  OTU  in the $j$th biological sample.
We  use the  notation
$n^j=\sum_{i=1}^{I} n_{i,j} $, $n_i=\sum_{j=1}^{J} n_{i,j}$,
$\bsigma=(\sigma_1,\ldots,\sigma_I)$, $\bY = (\bY^j,j=1,\ldots,J)$ and $\bQ = (Q_{i,j},1\leq i\leq I,1\leq j\leq J)$.
By  using  the    representation in  (\ref{SSS_ll})
we  introduce   the  latent  random   variables $\bT = (T_1,\ldots,T_J)$  and  rewrite  the posterior distribution of $(\bsigma,\bQ)$ :
\begin{align}
p(\bsigma,\bQ|\bn)\propto &
\left(\prod_{j=1}^J \prod_{i=1}^I \left(\sigma_iQ_{i,j}^{+2}\right)^{n_{i,j}} \right)\times \prod_{j=1}^J\left(\sum_{i=1}^I\sigma_i Q_{i,j}^{+2}\right)^{-n^j}\times\pi(\bsigma,\bQ)
\\
\propto &
\int \pi(\bsigma,\bQ)\prod_{j=1}^J \left\{\left(\prod_{i=1}^I \left(\sigma_iQ_{i,j}^{+2}\right)^{n_{i,j}}\right)
\frac{T_j^{n^j-1}\exp\left(-T_j\sum_i\sigma_i Q_{i,j}^{+2}\right) }{\Gamma(n^{j})}\right\}d\bT,
\end{align}
where    $\pi$  is  the  prior.
 In order to   obtain  approximate  $(\bsigma,\bQ)$   sampling
we  specify  a  Gibbs sampler  for $(\bsigma,\bQ,\bT)$   with  target  distribution
\begin{equation}
\label{post.s.q.T}
\begin{aligned}
p(\bsigma,\bQ,\bT|\bn)\propto &
\pi(\bsigma,\bQ)\prod_{j=1}^J \left\{\left(\prod_{i=1}^I \left(\sigma_iQ_{i,j}^{+2}\right)^{n_{i,j}}\right)
\frac{T_j^{n^{j}-1}\exp\left(-T_j\sum_i\sigma_i Q_{i,j}^{+2}\right) }{\Gamma(n^{j})}\right\}.
\end{aligned}
\end{equation}
The sampler iterates the following steps:

 [Step 1] Sample $T_j$ independently,  one  for  each biological sample    $j=1,\ldots,J$,
\begin{align}
T_j|\bQ,\bsigma,\bn \sim \text{Gamma}(n^{j},\sum_i\sigma_iQ_{i,j}^{+2}).\nonumber
\end{align}

  [Step 2] Sample $\bQ_{i}$  independently, one  for  each  OTU  $i=1,\ldots,I$.
The  conditional density of  $\bQ_i=(Q_{i,1}\ldots Q_{i,J})$    given  $\bsigma,\bT,\bn$  is  log-concave, and the random vectors
$\bQ_i$,  $i=1,\ldots, I$,  given $\bsigma,\bT,\bn$  are conditionally  independent.

 We simulate, for $j=1,\ldots,J$, from
\begin{align}
p(Q_{i,j}|\bQ_{i,-j},\bsigma,\bT,\bn)\propto Q_{i,j}^{+2n_{i,j}}\times
\exp\left(-T_j\sigma_i Q_{i,j}^{+2}\right)\times %
\exp\left(-\frac{(Q_{i,j}-\mu_{i,j})^2}{2s_{j}^2}\right),
\label{post.q.ind}
\end{align}
where $\bQ_{i,-j}=(Q_{i,1},\ldots,Q_{i,j-1},Q_{i,j+1},\ldots,Q_{i,J})$, $\mu_{i,j}=E[Q_{i,j}|\bQ_{i,-j}]$, $s_{j}^2=\var[Q_{i,j}|\bQ_{i,-j}]$,
  with the  proviso $0^0=1$. 
Since  $\bQ_i$  is  a  multivariate normal, both $\mu_{i,j}$  and  $s_j$  have  simple  closed form expressions.

When $n_{i,j}=0$ the density  in (\ref{post.q.ind}) reduces to a mixture of truncated normals:
\begin{align*}
&(1-p_1)N(Q_{i,j};\frac{\mu_{i,j}}{\Delta_{i,j}},\frac{s_{j}^2}{\Delta_{i,j}})I(Q_{i,j}> 0) +
 p_1N(Q_{i,j};\mu_{i,j},s_{j}^2)I(Q_{i,j}\leq 0),
\\
&p_1 = \frac{\Phi(0;\mu_{i,j},s_{j}^2)N(0;\frac{\mu_{i,j}}{\Delta_{i,j}},\frac{s_{j}^2}{\Delta_{i,j}})}{\Phi(0;\mu_{i,j},s_{j}^2)N(0;\frac{\mu_{i,j}}{\Delta_{i,j}},\frac{s_{j}^2}{\Delta_{i,j}}) + N(0;\mu_{i,j},s_{j}^2)\left(1-\Phi(0;\frac{\mu_{i,j}}{\Delta_{i,j}},\frac{s_{j}^2}{\Delta_{i,j}})\right)},
\end{align*}
and $\Delta_{i,j}=1+2\sigma_iT_js_{j}^2$.
 Here $N(\cdot;\mu,s^2)$ and $\Phi(\cdot;\mu,s^2)$ are  the density  and cumulative  density functions of a
normal   variable  with  mean  $\mu$  and  variance $s^2$.

When $n_{i,j}>0$  the  density $p[Q_{i,j}|\bQ_{i,-j},\bsigma,\bT,\bn]$   remains log-concave, and  the  support  becomes $(0,+\infty)$. We update  $Q_{i,j}$ using  a Metropolis-Hastings step with  proposal  identical to the
  Laplace approximation $N(\muhat_{i,j},\shat_{i,j}^2)$ of the density in (\ref{post.q.ind}), 
  \small
  \begin{align}
  \tiny
  \muhat_{i,j} &= \frac{\mu_{i,j}/s_{j}^2+\sqrt{\mu_{i,j}^2/s_{j}^4+8n_{i,j}(2\sigma_iT_j+1/s_{j}^2)}}{2(2\sigma_iT_j+1/s_{j}^2)},&
  \shat_{i,j}^2 &= \left(\frac{2n_{i,j}}{\muhat_{i,j}^2}+2T_j\sigma_i+\frac{1}{s_{j}^2}\right)^{-1}.
   \label{approx.para}
 \end{align}
 \normalsize
  Here $\muhat_{i,j}$ maximizes the density (\ref{post.q.ind}), and $\shat_{i,j}^2$ is obtained from the  second derivative of the log-density at $\muhat_{i,j}$.
  We  found   the approximation  accurate. In Supplementary Document S4 we  provide  bounds of  the  total variation distance  between  the  target   (\ref{post.q.ind})  and  the    approximation (\ref{approx.para}). When $n_{i,j}$ increases, the bound of the total variation decreases to  zero. See also Figure S1 in the Supplementary Document.

 [Step 3] Sample $\sigma_i$ independently, one for each OTU $i=1,\ldots,I$, from the density
$
p(\sigma_i|\bQ,\bT,\bn) \propto \pi(\sigma_i)\sigma_i^{n_i}\exp(-\sigma_i\sum_{j=1}^JT_jQ_{i,j}^{+2}).
$
The $\sigma_i$'s are a priori independent $\text{Beta}(\alpha/I,1/2-\alpha/I)$ variables.
We  use   piecewise constant bounds  for   $\sigma \rightarrow \exp(-\sigma_i\sum_{j=1}^JT_jQ_{i,j}^{+2})$,   $\sigma \in [0,1]$ and  an accept/reject  step  to  sample  from
 $p(\sigma_i|\bQ,\bT,\bn)$.

 We    now  consider  inference  on  $\bSigma$  using  the prior on $\bY$    in subsection \ref{third.subsec}. The  goal  is  to  generate  approximate  samples  of  $\bY$  from the  posterior. We exploit the identity of the conditional distributions of $\bY $ given $(\bsigma,\bT,\bQ,\bn)$  and $\bQ$.   In order to sample  $\bY$ from the posterior we can therefore directly  apply  the MCMC  transitions  in \cite{dunsonfactor},  with $\bQ$  replacing  the observable  variables  in their  work.

\subsection{Self-consistent estimates of biological samples' similarity}

We discuss an EM-type algorithm to  estimate  the correlation matrix   $\bS$ of the   vectors $(Q_{i,1},\ldots,Q_{i,J})$, $i=1,\ldots,I$.
Under our  construction in subsection \ref{third.subsec}, we interpret $\bS$ as the normalized version of Gram matrix $(\phi(j,j'))_{j,j'\in\Jsc}$ between biological samples. In this subsection we describe  an  alternative  estimating procedure,  distinct  from  the  Gibbs  sampler,  which does  not  require  tuning  of  the  prior  probability model.
The  algorithm  can be  used  for MCMC initialization  and for exploratory data analyses.
 It assumes  that the observed OTU abundances are representative of the  microbial distributions, i.e. $P^j=(n_{1,j}/n^j,\ldots,n_{I,j}/n^j)$.
Under this  assumption, for  each  biological sample $j$,
  \begin{align}
\sigma_iQ_{i,j}^{+2} \times  \mathbb I(n_{i,j}>0)\propto n_{i,j},~ i=1,\ldots,I,  \nonumber \\
\text{and }  Q_{i,j}\le 0 \;\text{ when }  \;n_{i,j}=0.
\label{obs.inf}
  \end{align}
For $\sigma_i$,  $i=1,\ldots,I$, we use a moment  estimate $\sigmahat_i=(1/J)\sum_j \left(n_{i',j}/\sum_{i\neq i'}n_{i,j}\right)$.
The   procedure uses  these   estimates  and   at  iteration $t+1$ generates  the  following  results:

\noindent [{\bf Expectation}] Impute   repeatedly $\bQ$,  $\ell=1,\ldots,D$ times,    consistently    with  the  constraints (\ref{obs.inf})  and using a $N(0,\bSigma_t)$ joint distribution.
Here $\bSigma_t$ is the estimate of  $\bSigma$, the covariance  matrix of $(Q_{i,1},\ldots,Q_{i,J})$,   after  the $t$-th iteration.
   For  each  replicate   $\ell=1,\ldots,D$, we
 fix  $Q_{i,j}^\ell$  for all  $(i,j)$ pairs  with strictly positive  $n_{i,j}$ counts at  $\sqrt{n_{i,j}/\sigmahat_i}$ and sample jointly, conditional on these values,  negative $Q_{i,j}^\ell$   values  for   the  remaining  $(i,j)$ pairs  with   $n_{i,j}=0$.
We use these   $Q_{i,j}^\ell$  values  to approximate  $\Lsc(\bSigma)$, the full data log-likelihood, our target   function  as  in any  other  EM  algorithm.

\noindent [{\bf Maximization}] Set $\bSigma_{t+1}$ equal  to  the empirical covariance matrix 
  of the  $(Q_{i,1}^{\ell},\ldots,Q_{i,J}^{\ell})$ vectors,  thus maximizing the $\mathcal{L}(\bSigma)$ approximation.

We  iterate  until  convergence  of  $\bSigma_{t}$.   Then, after  the  last  iteration,   the
inferred   covariance  matrix of $(Q_{i,1},\ldots,Q_{i,J})$  directly  identifies  an  estimate  of  $\bS$.
We evaluated  the  algorithm using  in-silico datasets  from the  simulation  study  in Section \ref{sec.sim}.
Overall      it  generates  estimates  that  are  slightly  less  accurate  compared to  posterior  estimation  based on    MCMC  simulations.
We use      the  datasets  considered  in Figure \ref{sim.res}(a), with number of factors fixed at three and $n^j$ at 100,000,   for a  representative   example.
In this  case the  average
RV-coefficient between   the  true  $\bS$  and  the   estimated matrix is 0.93  for  the  EM-type algorithm and   0.95   for  posterior  simulations.   
In our work the  described    procedure   reduced  the computing  time  to  approximately  10\% compared  to  the  Gibbs  sampler.
More details  on this procedure  are  provided   in  the Supplementary Document S5.

\section{Visualizing  uncertainty in ordination plots}
\label{sec.statis}
Ordination methods such as Multidimensional Scaling of ecological distances or Canonical Correspondence Analysis are central in microbiome research.
 Given posterior samples of the model parameters, we use a procedure to plot credible regions in  visualizations such as Fig \ref{sim.res}(f).
 The  methods that we  consider here are all related to PCA  and use
 the normalized Gram matrix $\bS$ between biological samples. We recall that in our model $\bS$  is the correlation matrix of $(Q_{i,1},\ldots,Q_{i,J})$.
  Based on  a single posterior instance of $\bS$, we can visualize  biological samples in a lower dimensional space through PCA,  with
each biological sample projected once.
   Naively, one could think that simply overlaying projections of  the principal component loadings generated from different posterior samples of $\bS$ on the same graph would show the variability of the projections.
   However, these super-impositions could be spurious if we carry out PCA for each $\bS$  sample  separately. One possible problem is principal component (PC) switching, when two PCs have similar eigenvalues.
Another problem is the ambiguity of signs in PCA, which
would lead to random signs of the loadings that result in
symmetric  groups of  projections of  the  same  biological sample at different sides of the axes.
More generally PCA projections from different posterior samples of $\bS$ are difficult to compare, as the different lower dimensional spaces
are not aligned. 

We  alternatively identify a consensus lower dimensional  space  for all posterior samples of $\bS$
 \citep{STATIS,ACT,distatis}.
We   list  the three  main  steps  used  to  visualize   the variability of $\bS$.
\begin{enumerate}
\item Identify a normalized Gram matrix $\bS_0$ that best summarizes $K$ posterior samples of normalized Gram matrix $\bS_1,\ldots,\bS_K$. One simple criterion is to minimize $L_2$ loss element-wise. This leads to $\bS_0=(\sum_i \bS_i)/K$. Alternatively, we can define $\bS_0$ as the normalized Gram matrix that maximizes similarity with $\bS_1,\ldots,\bS_K$.
One possible similarity metric between two symmetric square matrices $\bA$ and $\bB$ is the RV-coefficient \citep{rvcoef}, $\text{RV}(\bA,\bB) = \text{Tr}(\bA\bB)/\sqrt{\text{Tr}(\bA\bA)\text{Tr}(\bB\bB)}$.
We  refer to  \cite{rvreview} for  a discussion on RV-coefficients.

\item Identify the lower dimensional consensus space $V$ based on $\bS_0$. Assume we want $\dim(V)=2$; the basis of $V$ will be the orthonormal eigenvectors $\bv_1$ and $\bv_2$ of $\bS_0$ corresponding to the  largest eigenvalues $\lambda_1$ and $\lambda_2$. The configuration of all biological samples in $V$ is visualized by projecting rows of $\bS_0$ onto $V$:
$
(\bpsi^0_1,\bpsi^0_2) = \bS_0(\bv_1\lambda_1^{-1/2},\bv_2\lambda_2^{-1/2}).
$
As in a standard PCA, this configuration best approximates the normalized Gram matrix in the $L_2$ sense:
$
(\bpsi^0_1,\bpsi^0_2)=\argmin_{\langle\bpsi_1,\bpsi_2\rangle=0}\|\bS_0-(\bpsi_1,\bpsi_2)(\bpsi_1,\bpsi_2)'\|^2.
$

\item Project the rows of posterior sample $\bS_k$ onto $V$ by
$
(\bpsi^k_{1},\bpsi^k_{2}) = \bS_k(\bv_1\lambda_1^{-1/2},\bv_2\lambda_2^{-1/2}).
$
Overlaying all the $\bpsi^k$  displays  uncertainty  of $\bS$  in the same linear subspace.
Posterior variability of the  biological samples'  projections   is visualized   in $V$  by plotting  each  row  of the  matrices  $(\bpsi^k_{1},\bpsi^{k}_2)$,  $k=1,\ldots, K$, in the same figure.
 A contour plot is produced for each biological sample  (see for example  Fig \ref{sim.res}(f)) to  facilitate  visualization  of  the  posterior   variability  of its position in the consensus space $V$.
\end{enumerate}

\section{Simulation Study}
\label{sec.sim}  
In this section,
we evaluate  the  procedure  described in  Section \ref{sec.post} and explore whether the shrinkage prior  allows us  to infer  the  number of  factors and the normalized Gram matrix between biological samples $\bS$.
 We also  consider   the estimates $E(P^j | \bn)$ obtained  with our  joint  model,
one  for  each  biological sample $j$, and compare their precision
   with  the empirical estimator. Throughout the section, we assumed the number of factors is $m=10$ when running the posterior simulations.

We first   defined   a scenario  with distributions  $P^j$
 generated  from the prior (\ref{rmeasure}), with $I=68$ OTUs and  $J=22$ biological samples.
The true number of factors is $m_0$, and for biological samples  $j=1,\ldots, m_0/2$,  the vector  $\bY^j=(Y_{l,j},1\leq l\leq m_0)$ has  elements  $l=m_0/2+1,\ldots, m_0$ equal to  zero,
 while  symmetrically,  for    $j=J/2+1,\ldots, J$, the   vectors  $\bY^j$  have  the elements
  $l=1,\ldots,m_0/2$ equal to zero.
 The underlying normalized Gram matrix $\bS$ is therefore  block-diagonal.
 After generating the   distributions  $P^j$,
 we  sampled   with  fixed   total counts ($n^j$)  per biological sample   $n^j$= 1,000.
 We produced 50 replicates with $m_0=$3, 6, and 9.
In  our  simulations the non-zero components  $Y_{l,j}$'s are   independent  standard normal.

\begin{figure}
\centering
\includegraphics[scale=0.4]{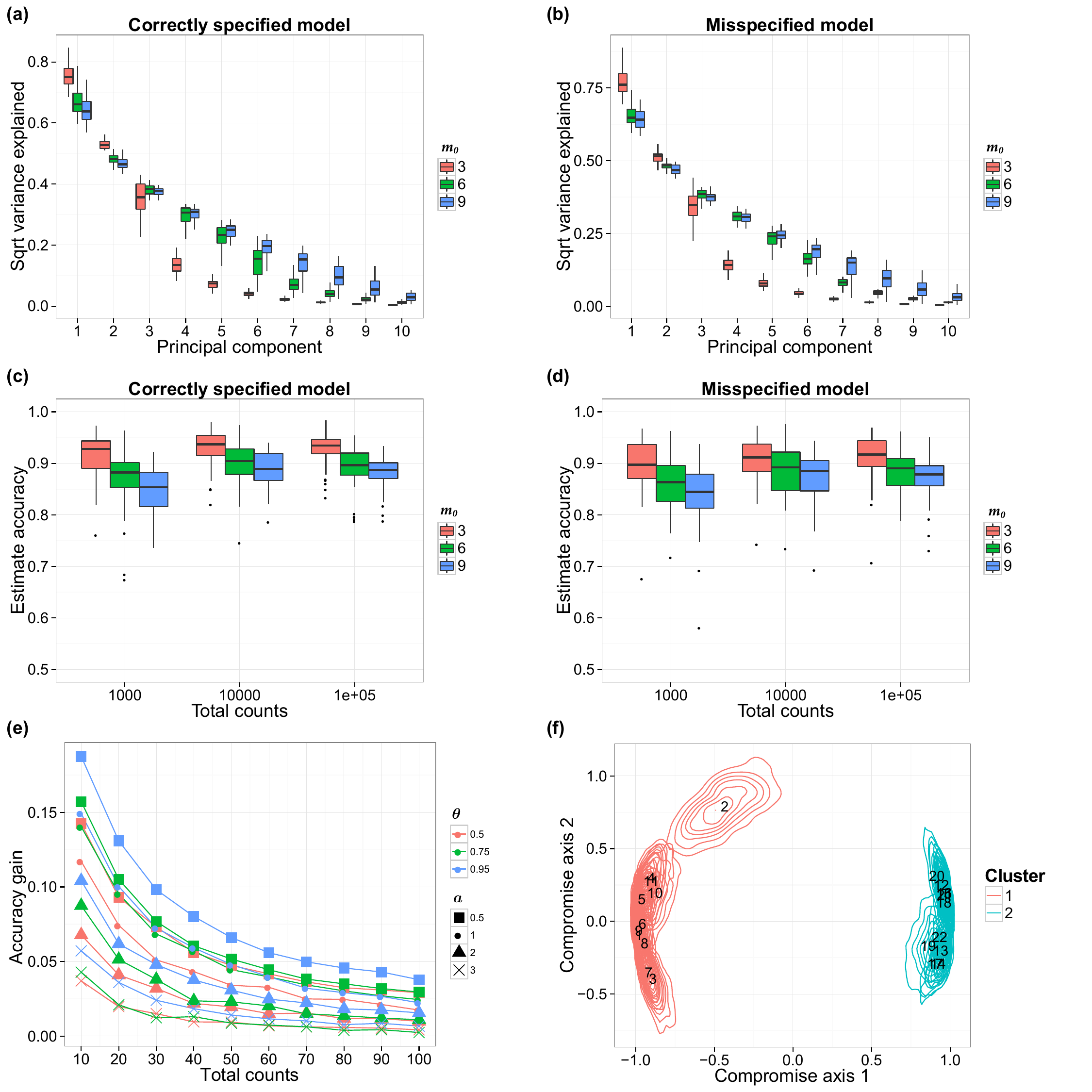}
\caption{\footnotesize{(\textbf{a-b})  Estimated proportion of variability captured by the first 10 PCs.
Each box-plot here shows the variability of the estimated proportion across 50 simulation replicates.
We show the results when the  data are generated from the prior (Panel \textbf{a}) and from the model in (\ref{mis.model}) with $a=1$ (Panel \textbf{b}).
(\textbf{c-d}) Accuracy of  the correlation matrix estimates $\bShat$.
The box-plots  show the variability of the accuracy in 50 simulation replicates, with  data generated from the prior (Panel \textbf{c}) and  from model   (\ref{mis.model}) with $a=1$  (Panel \textbf{d}).
 We  vary  the true number of factors $m_0$ (colors) and  $n^j$  and   show  the corresponding  accuracy  variations.
 (\textbf{e}) Comparison between  Bayesian estimates of the underlying microbial  distributions $P^j$ and the empirical estimates.
 We consider  the average    total variation  difference, averaging   across all $J$ biological samples.
  Each curve shows the relationship between $n^j$ and average accuracy gain.
  We set $m_0=3$ and    the parameter $a$  varies from 0.5 to 3 (shapes).  The   similarity  parameter   $\theta$  is  equal to  0.5, 0.75  or 0.95 (colors).
  (\textbf{f}) PCoA plot with confidence regions. We visualize  the confidence regions using  the  method in Section \ref{sec.statis}.
  Each contour  illustrates  the uncertainty   of  a  single   biological sample's position. Colors indicate  cluster  membership and annotated numbers are biological samples' IDs.}}
\label{sim.res}
\end{figure}

 We  use  PCA-type  summaries for  the  posterior samples of $\bY$
 generated   from    $p(\bY|\bn)$.
 Computations  are  based  on the  $J\times J$ normalized Gram  matrix $\bS$.
 At each   MCMC  iteration  we generate approximate  samples  $\bY$  from the
     posterior, compute  $\bS$  by normalizing the Gram matrix  $\bY'\bY$,    and  operate  standard  spectral decomposition on $\bS$.
        This  allows  us  to  estimate  the   ranked  eigenvalues, i.e. the principal components' variance of  our  $\bQ$  latent vectors (after normalization), by  averaging  over the  MCMC  iterations.
 Figure \ref{sim.res}(a)
   shows  the variability  captured   by the first 10 principal components,
   with  the  box-plots   illustrating posterior means' variability across our 50 replicates.
        %
  The proportion of variability
  associated  to  each  {\it principal  component}  decreases  rapidly  after the {\it true}  number of   factors  $m_0=3,6,9$.
This suggests that
  the  shrinkage  model \citep{dunsonfactor} tends  to   produce posterior   distributions  for
our   $\bY$  latent  variables   that  concentrates  around  a linear  subspace. 

 Figure \ref{sim.res}(c) illustrates the  accuracy  of the estimated normalized Gram matrix $\bShat$ with $n^j$ equal to 1,000, 10,000, and 100,000.
  We estimated the unknown $J\times J$ normalized Gram matrix $\bS$ with the posterior mean of the normalized Gram matrix, which we approximate by averaging over  MCMC iterations.
  We  summarized  the  accuracy using  the RV coefficient between $\bShat$ and $\bS$,  see \cite{rvcoef} for  a  discussion  on  this metric.
The box-plots illustrate variability of estimates' accuracy across 50 simulation replicates. As  expected,  when the total counts per sample   increases  from  10,000  to  100,000, we  only observe   limited   gain  in   accuracy.
   Indeed  the  overall  number  of  observed   OTUs  with  positive  counts  per  biological sample
           remains  comparable, with  expected  values  equal to 30  and  33   when  the  total  counts  per  biological sample  are  fixed  at  10,000  and  100,000 respectively. We also note that when $m_0$ increases, the accuracy decreases.

We investigate   interpretability of our model  by  using
       distributions  $P^j$ generated from a probability  model that  slightly differs   from the prior.
 More  precisely,   the   $i$th
 random  weight   in   $P^j$,
   conditionally   on $\bY$ and $\bX$,  is  defined  proportional
   to     a   monotone function of $\langle \bX_i,\bY^j \rangle^+$.  We  considered for  example
\begin{equation}
P^{j}(A) = \frac{\sum_i \sigma_i\langle \bX_i,\bY^j\rangle^{+a}I_{Z_i}(A)}{\sum_i \sigma_i\langle \bX_i,\bY^j\rangle^{+a}},\;\;\;a>0.
\label{mis.model}
\end{equation}
When  the monotone  function  is  quadratic  the  probability  model  becomes  identical  to  our  prior.
In  Figure \ref{sim.res}(b) and Figure \ref{sim.res}(d)  we  used  model (\ref{mis.model})
with  $a=1$ to  generate  datasets.  We  repeated  the same  simulation  study  summarized  in the  previous  paragraphs.

We evaluated  the effectiveness of borrowing information  across biological samples for  estimating
the  vectors $P^j$.
The  accuracy  metric  that  we  used  is  the total variation distance.
 We compared the Bayesian estimator $E(P^j|\bn)$ and
 the   empirical  estimator  $\tilde  P^j$ which  assigns  mass  $n_{i,j}/n^j$ to  the  $i^{th}$  OTUs. The advantage of pooling information 
  varies   with
   the similarity between biological samples. To  reflect  this, we  generated $P^j$ with
  non-zero components of $\bY$ sampled  from  a  zero  mean  multivariate  normal
   with $ \cov(Y_{l,j},Y_{l,j'})$  equal  to $ \theta$.
    We considered the case when $P^j$ is generated either from our prior or model (\ref{mis.model}) with $a=0.5,1,3$.
    In addition, we considered $\theta=0.5,0.75,0.95$, $I=68,J=22$, and $m_0=3$, while $n^j$ varies from 10 to 100.

 The  results  are  summarized  in Figure \ref{sim.res}(e)  which  shows  the average difference in total variation,
 contrasting  the  Bayesian  and  empirical  estimators.
  The results, both
 when  the  model is  correctly specified,  and  when mis-specified,   quantify the  advantages  in  using  a  
 joint Bayesian model.

We  complete  this section  with  one  illustration of  the method in Section \ref{sec.statis}.
 We simulate a dataset with  two clusters by
   generating $Y_{l,j}$ for $l=1,\ldots,m_0$ from $N(-3,1)$ when $j=1,\ldots,J/2$ and from $N(3,1)$ when $j=J/2+1,\ldots,J$.
   All $Y_{l,j}$  are different  from zero.
     We expected   a  low  $n^j$  to be  sufficient  for  detecting  the    clusters.
     We sampled $P^j$ from the prior and set $J=22$, $I=68$, $m_0=3$, and $n^j=100$.
     The PC plot  and  the biological sample specific credible regions are shown in Figure \ref{sim.res}(f).
     In  the  PC   plot   the  two  clusters  are  illustrated  with  different  colors. In this  simulation exercise the posterior credible regions leave
     little  ambiguity  both  on  the presence  of  clusters  and also  on   samples-specific  cluster  membership. To compare this with the Principal Coordinates Analysis (PCoA) method used in microbiome studies, we plot the ordination results using PCoA based on the Bray-Curtis dissimilarity matrix derived from the empirical microbial distributions (See Figure S3). 
     We can see that the  PCoA point estimate  is similar to the centroids identified  by the  proposed Bayesian ordination method.

\section{Application to microbiome datasets}

In this section, we  apply our Bayesian analysis to  two microbiome datasets. We show that our method gives results that are consistent with previous studies, and we show our novel visualization of uncertainty in ordination plots. We start with  the Global Patterns data \citep{gp_paper} where human-derived and environmental biological samples are included. We then considered data on the vaginal microbiome \citep{ravel_vaginal}.
 
\subsection{Global Patterns dataset}

The Global Patterns dataset includes 26 biological samples derived from both human and environmental specimens.
 There are a total of 19,216 OTUs, and the average total counts per biological sample is larger than 100,000.  
 We collapsed all taxa OTUs to the genus  level---a  standard operation in microbiome studies---and  yielded 996 distinct genera. We treated these genera as OTUs' and fit our model to this collapsed dataset. 
 We ran one MCMC chain for 50,000 iterations and recorded posterior samples every 10 iterations.

\begin{figure}[htbp]
\centering
\includegraphics[scale=0.45]{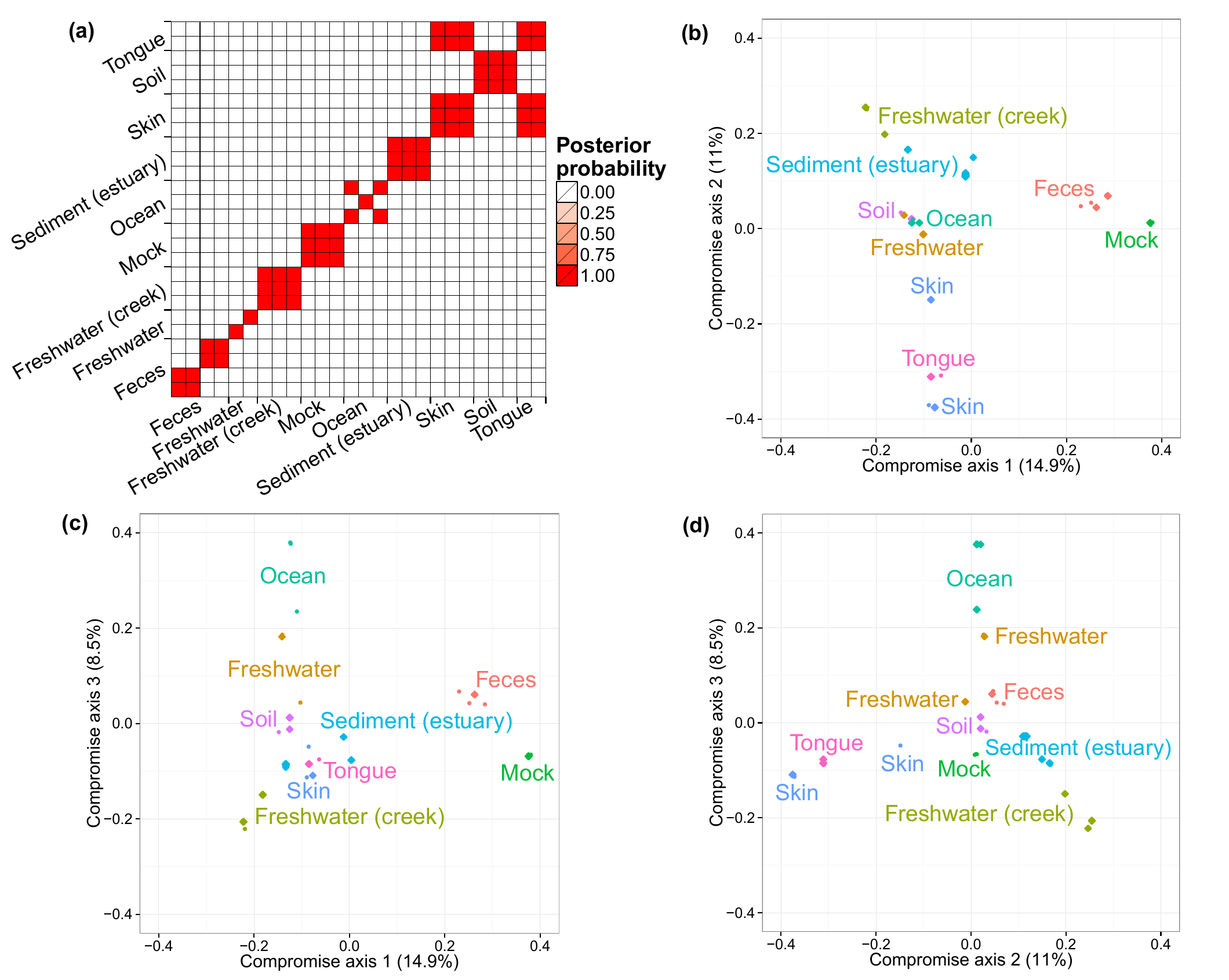}
\caption{\footnotesize{(\textbf{a}) Posterior Probability of each pair of biological samples $(j,j')$ being clustered together. 
The labels on axes indicate the environment of  origin for  each  biological sample.
 (\textbf{b-d}) Ordination plots of biological samples and 95\% posterior credible regions. 
We illustrate the first three compromise axes with three panels. 
Panel  (b)  plots  projections  on  the  first  and  second axes.
Panel  (c)  plots  projections  on  the  first  and   third  axes.
Panel  (d)  plots  projections  on  the  second  and  third axes.
The percentages on the three axes are the ratios of the corresponding $\bS_0$ eigenvalues and the trace of the matrix. The credible regions for some biological samples are so small that appears as single points. Colors and annotated text indicate the environments.}}
\label{gp.fig}
\end{figure}

We first performed a cluster analysis of biological samples based on their microbial compositions. For each posterior sample of the model parameters, we computed $P^j$ for $j = 1, \ldots, J$ and calculated the Bray-Curtis dissimilarity matrix between biological samples. 
We then clustered the biological samples using this dissimilarity matrix with Partitioning Among
Medoids (PAM) \citep{hastie2003pam}. By averaging over the MCMC iterations for the clustering results from
each dissimilarity matrix, we obtained the posterior probability of two biological samples
being clustered together.
Figure \ref{gp.fig}(a) illustrates the  clustering probabilities.
 We can see that biological samples belonging to a specific specimen type are tightly clustered together   while   
 different specimens   tend  to  define  separate  clusters. 
 This is consistent with the conclusion in \cite{gp_paper}, where the authors suggest, that within specimen
 microbiome   variations  are    limited  when  compared   to  variations 
     across    
  specimen types. 
 We also observed that biological samples from the skin are clustered with those from the tongue. 
 This is  to some  extent  an  expected  result, 
   because  both  specimens are  derived from humans, and 
   because  the skin microbiome   has  often   OTUs   frequencies   comparable  to  other   body sites \citep{grice2011skin}.

We then visualized the   biological samples using ordination plots   and   applying the method described in Section \ref{sec.statis}. 
We fixed the dimension of the consensus space $V$ at three. 
We plotted  all biological samples' projections  onto  $V$ along with contours  to  visualize  their  posterior variability.
The  results  are shown in Figure \ref{gp.fig}(b-d).
We observe  a clear separation between human-derived (tongue, skin, and feces) biological samples and biological samples from free environments.
 This separation is mostly identified by the first two compromise axes. 
 The  third  axis  defines  a  saline/non-saline samples  separation. 
 Biological samples derived from saline environment (e.g. Ocean) are well separated   when  projected on  this axis from those derived from non-saline environment (e.g. Creek freshwater). 
  We  observed  small  95\% credible regions for  all biological samples projections. This  low level  of  uncertainty captured by  the small credible regions in   Figure \ref{gp.fig}(b-d) is mainly    explained  by  the  large total counts $n^j$   for  all biological  samples. Finally, to compare the ordination results  to those given by  standard  methods used in microbiome studies, we  generated ordination results using  PCoA. Figure S4  shows that  the   relative  positions  of  different types of biological samples in  PCoA plots  and in  the Bayesian ordination plots are  similar.
  
\subsection{The Vaginal Microbiome}
We also consider a  dataset  previously presented  in \cite{ravel_vaginal}  which contains  a  larger  number of  biological samples (900) and  a simpler bacterial community structure. 
These biological samples are derived from 54 healthy women. 
Multiple biological samples are taken from each individual, ranging from  one  to 32 biological samples per  individual.
 Each  woman has  been  classified,  before our  microbiome  sequencing   data  were   generated,  into     vaginal community state subtypes (CST). 
 This dataset contains only species level taxonomic information, and we filtered OTUs by occurrence.
  We only retain species  with  more than five reads in at least 10\% of biological samples. 
   This filtering resulted in 31 distinct OTUs. We ran one MCMC chain with 50,000 iterations.

\begin{figure}[htbp]
\centering
\includegraphics[scale=0.32]{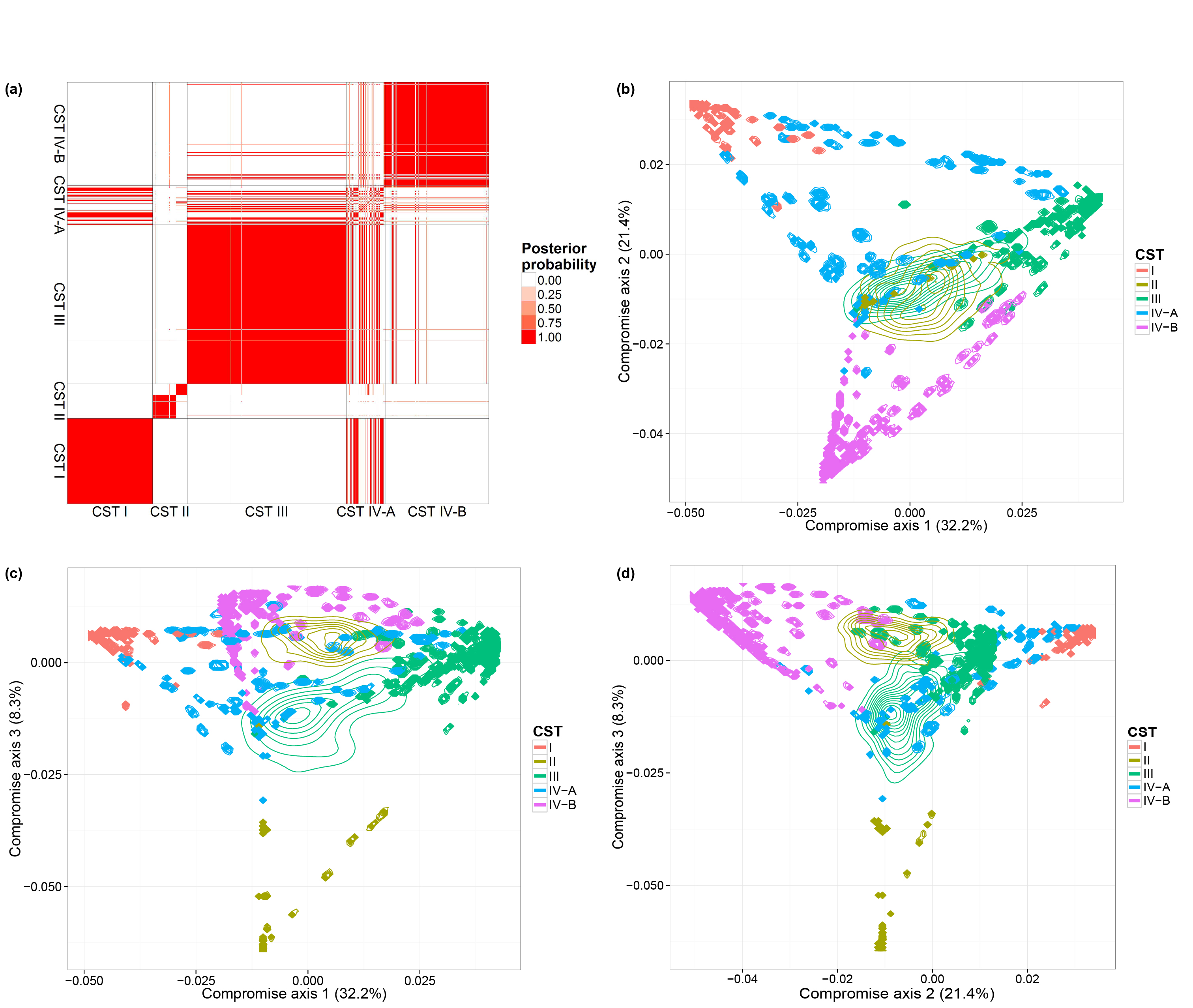}
\caption{\footnotesize{(\textbf{a}) Posterior Probability of each pair of biological samples $(j,j')$ being clustered together. 
The labels on axes indicate the    CST  for  each  biological sample.
 (\textbf{b-d}) Ordination plots of biological samples and  posterior credible regions. 
 We  illustrate  the first three compromise axes with three panels. The percentages on the three  axes  are  the ratios of  the  corresponding  $\bS_0$  eigenvalues  and  the trace of the matrix.
 Colors and indicate CSTs.}}
\label{ravel.fig}
\end{figure}

We performed the same analyses as in the previous subsection. 
The results are shown in Figure \ref{ravel.fig}.
   Clustering probabilities  indicate strong within CST similarity (panel a). 
 There  is  one     exception,   
  CST IV-A  samples,     in  some  cases,       presenting low  levels  of    similarities    when  compared to  each  other
       and   tend   to  cluster  with 
  CST I, CST III, and CST IV-B  samples. 
 This is because CST IV-A is characterized as a highly  heterogeneous  subtype \citep{ravel_vaginal}. 
 The ordination plots are consistent with the discoveries in \cite{ravel_vaginal}. 
 A tetrahedron shape is recovered, and CST I, II, III, IV-B occupy the four vertices.
  CST II is well separated from other CSTs by the third axis. This pattern is  similar  to  the  one   observed in the  plots generated using PCoA  (Figure S5).
  We also observed a sub-clustering in CST II   which  has  not  been  detected  and  discussed  in  \cite{ravel_vaginal}.
 This  difference  in  the  results  can  be    due  to    distinct     clustering
   metrics  in the  analyses. 

Note that there are two biological samples with  large credible regions, indicating high uncertainty of the corresponding positions.
    This  uncertainty  propagates   on  their  cluster  membership.     
 Both  biological samples have  small total counts  compared to the others.
  The lack of precision when using biological samples with small sequencing depth  leads to high uncertainty in ordination and  classification.
          It  is  therefore  important  to  account  for  uncertainty    in  the  validation  of   subgroups   biological  differences---in  our  case   CST  subtypes---based on      microbiome   profiling.
               Our  example  suggests  also  the  importance  of  uncertainty  summaries  when  microbiome  profiles  are  used  to  classify  samples.  
Uncertainty  summaries  allow us to retain all samples,  including  those with low counts,  without  the  risk  of    overinterpreting  the  estimated  locations   and  projections. 
   This also argues for the retention of raw counts in microbiome studies \citep{wastenotwantnot}. 
   By using raw counts, we can evaluate  the  uncertainty of   our  estimates  and exploit the information  and  statistical power carried by the full dataset; 
   whereas if we downsample  the data  we lose  information    and  increase  uncertainty on the projections.

It is ubiquitous to have       biological samples with   relevant     differences  in their    total    counts, and  in  some  cases  the  number  of  OTUs and  the  total number  of  reads can be  comparable.
 In this cases, the empirical estimates of microbial distributions  are not reliable, and an assessment of the uncertainty is necessary for downstream analyses. 
 The two biological samples with  low total counts in the vaginal microbiome dataset are  examples. 
For  biological  samples  with   a  scarce  amount of  data  our model      provides   measures  of  uncertainty   and  allows   uncertainty  visualizations  with  ordination  plots.

\section{Conclusion}

We propose a joint model for multinomial sampling of OTUs in multiple biological samples.
 We apply a  prior from Bayesian factor analysis  to estimate the similarity between biological samples, which is summarized by  a Gram matrix.
Simulation studies give evidence that this parameter is recovered by the Bayes estimate, and in particular, the inherent dimensionality of the latent factors is effectively  learned from the data.
The simulation  also  demonstrated that the analysis yields more accurate estimates of the microbial distributions by borrowing information across biological samples.

In addition, we provide a robust method to visualize the uncertainty in ecological ordinations, furnishing each point in the plot with a credible region.
Two published microbiome datasets were analyzed, and the results are consistent with previous findings.
The second analysis demonstrates that the level of uncertainty can vary across biological samples due to differences in sampling depth, which underlines the importance of modeling multinomial sampling variations coherently.
We believe our analysis will mitigate artifacts arising from rarefaction, thresholding of rare species, and other preprocessing steps.

There are several directions for development which are not explored here. We highlight the possibility of incorporating prior knowledge about the biological samples, such as the subject or group identifier in a clinical study. To achieve this, we can augment the latent factors $\bY^j$ by a vector of covariates $(b_1w^j_1,\dots,b_pw^j_p)$, whose coefficients $b$ could be given a normal prior, for example.
The posterior distribution of the coefficients could be used to infer the magnitude of  covariates'   effects.
A less straightforward extension involves moving away from the assumption of {\it a priori} exchangeability between OTUs to include prior information about phylogenetic or functional relationships between them.
 In our present analysis, these relationships  are not taken into account in the definition of the prior for microbial distributions. 

\section{Acknowledgements}

B. Ren is supported by National Science Foundation under Grant No. DMS-1042785. S. Favaro is supported by the European Research Council (ERC) through StG N-BNP 306406. L. Trippa has been  supported by the  Claudia Adams Barr Program in Innovative Basic Cancer Research. S. Holmes was supported by the NIH grant
R01AI112401.
We thank Persi Diaconis, Kris Sankaran and Lan  Huong Nguyen for helpful suggestions and improvements.

\begin{appendices}
\renewcommand{\thefigure}{S\arabic{figure}}
\renewcommand{\theequation}{S\arabic{equation}}
\renewcommand{\thesection}{S\arabic{section}}
\renewcommand{\thetable}{S\arabic{table}}
\setcounter{figure}{0}
\section{Approximating a Poisson Process using Beta random variables}
Consider approximating a Poisson process on $(0,1)$ with intensity $\nu(\sigma) = \alpha\sigma^{-1}(1-\sigma)^{-1/2}$ by a finite counting process formed by $n$ iid samples drawn from $\text{Beta}(\epsilon_n,1/2-\epsilon_n)$ where $\epsilon_n<1/2$. Denote the Poisson process as $N(t)$ and the approximating process as $N'_n(t)$, we first calculate the probability of having $m$ points in interval $(\delta,t]$, where $m\leq n$, $t<1$ and $0<\delta\ll 1$,
\begin{align*}
P\left[N((\delta,t])=m\right] =& \frac{\left[\int_{\delta}^t \alpha\sigma^{-1}(1-\sigma)^{-1/2}d\sigma\right]^m}{m!}\exp\left(-\int_{\delta}^t \alpha\sigma^{-1}(1-\sigma)^{-1/2}d\sigma\right),\\
P\left[N'_n((\delta,t])=m\right] =& \compn{n}{m}\left(\frac{1}{\text{Beta}(\epsilon_n,1/2-\epsilon_n)}\int_{\delta}^t \sigma^{-1+\epsilon_n}(1-\sigma)^{-1/2-\epsilon_n}d\sigma\right)^m\times\\
&~~\left(1-\frac{1}{\text{Beta}(\epsilon_n,1/2-\epsilon_n)}\int_{\delta}^t \sigma^{-1+\epsilon_n}(1-\sigma)^{-1/2-\epsilon_n}d\sigma\right)^{n-m}.
\end{align*}
The moment generating functions (MGFs) of $N((\delta,t])$ and $N'_n((\delta,t])$ are
\begin{align*}
M_{N}(\lambda) =& \exp\left[\left(e^{\lambda}-1\right)\int_{\delta}^t \alpha\sigma^{-1}(1-\sigma)^{-1/2}d\sigma\right],\\
M_{N'_n}(\lambda) =& \left[\frac{e^{\lambda}-1}{\text{Beta}(\epsilon_n,1/2-\epsilon_n)}\int_{\delta}^t \sigma^{-1+\epsilon_n}(1-\sigma)^{-1/2-\epsilon_n}d\sigma+1\right]^n.
\end{align*}
These two MGFs will be the same asymptotically if
\begin{equation}
\lim_{n\to\infty} \frac{n}{{\text{Beta}}(\epsilon_n,1/2-\epsilon_n)}\int_{\delta}^t \sigma^{-1+\epsilon_n}(1-\sigma)^{-1/2-\epsilon_n}d\sigma = \alpha\int_{\delta}^t \sigma^{-1}(1-\sigma)^{-1/2}d\sigma.
\label{cond.sigma}
\end{equation}
This will be satisfied when $\epsilon_n=\alpha/n$. Indeed, under this assumption, we have
$$
\lim_{n\to\infty}\frac{n\left(\sigma/(1-\sigma)\right)^{\epsilon_n}}{\text{Beta}(\epsilon_n,1/2-\epsilon_n)}=\alpha.
$$
In addition, since when $n$ is large enough, the map $n\mapsto \frac{n\left(\sigma/(1-\sigma)\right)^{\epsilon_n}}{\text{Beta}(\epsilon_n,1/2-\epsilon_n)}$ is a non-increasing function, by Lebesgue's monotone convergence theorem, we can establish the convergence of the left hand side of (\ref{cond.sigma}) to the right hand side. Using this result, we can prove the weak convergence of the finite dimension distribution: $(N'(\delta,t_1],\ldots,N'(\delta,t_n])\overset{d}{\to}(N(\delta,t_1],\ldots,N(\delta,t_n])$. This follows by a direct application of the multinomial theorem.

Now we need to verify the tightness condition, this is automatically satisfied as $N_n(t)'$ is a c\`adl\`ag process \citep{Vere} (Theorem 11.1. VII and Proposition 11.1. VIII, iv, Volume 2). Therefore we prove the weak convergence of the process $N'_n(t)$ to the Poisson process $N(t)$ when $n\to\infty$ and $\epsilon_n=\alpha/n$.

\section{Proof of Proposition 1}
\label{proof.prop1}
We use the
notation  $P^{j}(\cdot) = \frac{\sum_iI(Z_i \in \cdot)\sigma_iQ_{i,j}^{+2}}{\sum_i \sigma_i Q_{i,j}^{+2}}$ where $Q_{i,j}=\langle \bX_i,\bY^{j}\rangle$. Denote $ \left((Q_{i,j},Q_{i,j'}),i\geq 1\right)$ as $\bQ$.
The joint distribution of $(Q_{i,j},Q_{i,j'})$ is a multivariate normal with mean $\bzero$ and covariance  $\phi(j,j')$,
and  the vectors $(Q_{k,j},Q_{k,j'})$,  $k=1,2,\ldots,$  are  independent.  We derive an expression for the covariance

\begin{align*}
\cov[P^{j}(A),P^{j'}(A)] =& E[E[P^{j}(A)P^{j'}(A)|\sigma,\bQ]] - E[P^{j}(A)]E[P^{j'}(A)]\\
=&(G(A)-G^2(A))E\left[\frac{\sum_i \sigma_i^2Q_{i,j}^{+2}Q_{i,j'}^{+2}}{\sum_i \sigma_iQ_{i,j}^{+2}\sum_k \sigma_kQ_{k,j'}^{+2}}\right].
\end{align*}

Similarly, we can get the expression for the variance,
$$
\var[P^{j}(A)] = (G(A)-G^2(A))E\left[\frac{\sum_i \sigma_i^2Q_{i,j}^{+4}}{\sum_i \sigma_iQ_{i,j}^{+2}\sum_k \sigma_kQ_{k,j}^{+2}}\right].
$$

It follows that
$$
\corr[P^{j}(A),P^{j'}(A)] = E\left[\frac{\sum_i \sigma_i^2Q_{i,j}^{+2}Q_{i,j'}^{+2}}{\sum_i \sigma_iQ_{i,j}^{+2}\sum_k \sigma_kQ_{k,j'}^{+2}}\right]\mathlarger{ \times }
\left(E\left[\frac{\sum_i \sigma_i^2Q_{i,j}^{+4}}{\sum_i \sigma_iQ_{i,j}^{+2}\sum_k \sigma_kQ_{k,j}^{+2}}\right]\right)^{-1}.
$$
Therefore the correlation is independent of the set $A$.

\section{Proof of Proposition 2}

We follow the framework of proofs for Theorem 1 and Theorem 3 in \cite{DPP_proof}. Let $\Psc(\Zsc)$ be the set of all Borel probability measures defined on $(\Zsc,\Fsc)$ and $\Psc(\Zsc)^J$ the product space of $J$ $\Psc(\Zsc)$. Assume $\Theta\subset\Zsc$ is the support of $G$. To show the prior assigns strictly positive probability to the neighbourhood in Proposition 2, it is sufficient to show such neighbourhood contains certain subset-neighbourhoods with positive probability. As in \cite{DPP_proof}, we consider the subset-neighbourhoods $U$:
$$
U(G_1,\ldots,G_J,\{A_{i,j}\},\epsilon^*)=\prod_{i=1}^J \{F_i\in\Psc(\Theta):|F_i(A_{i,j})-G_i(A_{i,j})|<\epsilon^*,j=1,\ldots,m_i\},
$$
where $G_i$ is a probability measure absolutely continuous w.r.t. $G$ for $i=1,\ldots,J$, $A_{i,1},\ldots,A_{i,m_i}\subset \Theta$ are measurable sets with $G_{i}$-null boundary and $\epsilon^*>0$. The existence of such subset-neighbourhoods is proved in \cite{DPP_proof}. We then define sets $B_{\nu_{1,1}\ldots\nu_{m_J,J}}$ for each $\nu_{i,j}\in\{0,1\}$ as
$$
B_{\nu_{1,1}\ldots\nu_{m_J,J}} = \bigcap_{i=1}^J\bigcap_{j=1}^{m_i} A_{i,j}^{\nu_{i,j}},
$$
where $A_{i,j}^1=A_{i,j}$ and $A_{i,j}^0 = A_{i,j}^c$. Set
$$
J_{\nu}=\{\nu_{1,1}\ldots\nu_{m_J,J}:G(B_{\nu_{1,1},\ldots,\nu_{m_J,J}})>0\},
$$
and let $\Msc$ be a bijective mapping from $J_{\nu}$ to $\{0,\ldots,k\}$ where $k=|J_{\nu}|-1$. We can simplify the notation using $A_{\Msc(\nu)}=B_{\nu}$ for every $\nu\in J_{\nu}$. Define a vector $\bs_i=(w_{i,0},\ldots,w_{i,k})=(Q_i(A_0),\ldots,Q_i(A_k))$ that belongs to the $k-$simplex $\Delta_k$. Set
$$
B(\bs_i,\epsilon) = \{(w_0,\ldots,w_k)\in\Delta_k: |Q_i(A_j)-w_j|<\epsilon,j=0,\ldots,k\},
$$
where $\epsilon = 2^{-\sum_{i=1}^J{m_i}}\epsilon^{*}$. The derivation in \cite{DPP_proof} suggests a sufficient condition for assigning positive mass to $U(G_1,\ldots,G_J,\{A_{i,j}\},\epsilon^*)$ is
\begin{equation}
\Pi([P^i(A_0),\ldots,P^i(A_k)]\in B(\bs_i,\epsilon),i=1,\ldots,J)>0.
\label{equ.cond}
\end{equation}
Here $\Pi$ is the prior.

Now consider the following conditions
\begin{enumerate}
\item[C.1] $w_{i,l}-\epsilon_0 < \sigma_{l+1}Q_{l+1,i}^{+2} < w_{i,l}+\epsilon_0$ for $i=1,\ldots,J$ and $l=0,\ldots,k$.
\item[C.2] $0 < \sum_{l>k+1}\sigma_lQ_{l,i}^{+2} < \epsilon_0$.
\item[C.3] $Z_{l+1}\in A_l$ for $l=0,\ldots,k$.
\end{enumerate}
$\epsilon_0$ in the above conditions satisfies the following inequality
\begin{align*}
&\frac{w_{(i,l)}-\epsilon_0}{1+(k+2)\epsilon_0} \geq w_{(i,l)} - \epsilon\\
&\frac{w_{(i,l)}+2\epsilon_0}{1-(k+1)\epsilon_0} \leq w_{(i,l)} + \epsilon
\end{align*}
for $i=1,\ldots,J$ and $l=0,\ldots,k$. This system of inequalities can be satisfied when $k$ is large enough. If conditions $(C.1)$ to $(C.3)$ hold, it follows that $[P^i(A_0),\ldots,P^i(A_k)]\in B(\bs_i,\epsilon)$ for $i=1,\ldots,J$. Therefore, we have
\begin{align*}
&\Pi([P^i(A_0),\ldots,P^i(A_k)]\in B(\bs_i,\epsilon),i=1,\ldots,J) \geq\\
&\prod_{l=0}^k\Pi(w_{(i,l)}-\epsilon_0 < \sigma_{l+1}Q_{l+1,i}^{+2} < w_{(i,l)}+\epsilon_0, i=1,\ldots,J)\times\\
&\Pi(\sum_{l>k+1}\sigma_lQ_{l,i}^{+2}<\epsilon_0, i=1,\ldots,J)\times\\
&\prod_{l=0}^k\Pi(Z_{l+1}\in A_l)\times\Pi(Z_{l}\in \Zsc, l=k+2,\ldots).
\end{align*}

Since $(Q_{l,1},\ldots,Q_{l,J})$ are multivariate normal random vectors with strictly positive definite covariance matrix and $\sigma_l$ are always positive, the vector $(\sigma_{l+1}Q_{l+1,i}^{+2},i=1,\ldots,J)$ has full support on $\mathbb{R}^{+J}$ and will assign positive probability to any subset of the space. If follows that
$$
\Pi(w_{i,l}-\epsilon_0 < \sigma_{l+1}Q_{l+1,i}^{+2} < w_{i,l}+\epsilon_0, i=1,\ldots,J)>0\text{ for }l=0,\ldots,k.
$$

Using the Gamma process argument, we know $\sum_{l>k+1}\sigma_lQ_{l,i}^{+2}$ is the tail probability mass for a well-defined Gamma process and thus will always be positive and continuous for all $i$. It follows that
$$
\Pi(\sum_{l>k+1}\sigma_lQ_{l,i}^{+2}<\epsilon_0, i=1,\ldots,J)>0.
$$
Since $\Zsc$ is the topological support of $G$, it follows that
$P(Z_{i+1}\in A_i)>0$ and $P(Z_{i}\in \Zsc)=1$. Combining these facts, we prove that Equation (\ref{equ.cond}) holds.

\section{Total variation bound of Laplace approximate of \texorpdfstring{$p(Q_{i,j}|\bQ_{i,-j},\bsigma,\bT,\bn)$}{pQ}}
\label{tv.bound}
We consider the class of densities $g(x;k,\mu,s^2)$
$$
g(x;k,\mu,s^2) \propto I(x\geq0)x^{2k}f(x;\mu,s^2), k\in\mathbb{N}^+
$$
where $f(x;\mu,s^2)$ is the density function of $N(\mu,s^2)$. The Laplace approximation of $g(x;k,\mu,s^2)$ is written as $f(x;\muhat,\shat^2)$. Here $\muhat=\text{argmax}_{x}g(x;k,\mu,s^2)$ and $\shat^2 = -\left( \left(\partial^2 \log(g)/\partial x^2\right)|_{\muhat}\right)^{-1}$. We want to calculate the total variation distance between density $f(x;\muhat,\shat^2)$ and $g(x;k,\mu,s^2)$, denoted as $d_{TV}(f(x;\muhat,\shat^2), g(x;k,\mu,s^2))$.

Define class of functions $V(x;k,\mu)$ for $k\in\mathbb{N}^{+},\mu>0$:
$$
V(x;k,\mu) = \left\{
\begin{array}{cc}
2k\left[\log(x/\mu)-(x/\mu-1)+\frac{1}{2}(x/\mu-1)^2\right]&x>0\\
-\infty&x\leq 0
\end{array}\right.
$$
This function is non-decreasing and when $x=\mu$, $V(x;k,\mu)=0$, $dV/dx=0$ and $d^2V/dx^2=0$.

It follows that
$$
\log g(x;k,\mu,s^2) - \log f(x;\muhat,\shat^2) = V(x;k,\muhat) + a_0 + a_1x + a_2x^2.
$$
Moreover, since the $\muhat$ is the mode of both $g(x;k,\mu,s^2)$ and $f(x;\muhat,\shat^2)$, and the second derivative of $\log g(x;k,\mu,s^2)$ and $\log f(x;\muhat,\shat^2)$ are identical at $x=\muhat$, we can find that $a_1=a_2=0$. Hence,
$$
\log g(x;k,\mu,s^2) - \log f(x;\muhat,\shat^2) = V(x;k,\muhat) + a_0
$$
and $g(x;k,\mu,s^2) = \exp\left(V(x;k,\muhat)+a_0\right)f(x;\muhat,\shat^2)$.

Since $V(x;k,\muhat)$ is monotone increasing, the total variation distance between $g(x;k,\mu,s^2)$ and $f(x;\muhat,\shat^2)$ can be expressed as
\begin{align*}
d_{TV}(g(x;k,\mu,s^2),f(x;\muhat,\shat^2)) =& \int_{x_0}^{+\infty}\left[\exp\left(V(x;k,\muhat)+a_0\right)-1\right]f(x;\muhat,\shat^2)dx\\
=&\int_{-\infty}^{x_0}\left[1-\exp\left(V(x;k,\muhat)+a_0\right)\right]f(x;\muhat,\shat^2)dx
\end{align*}
where $V(x_0;k,\muhat) = -a_0$. If $a_0\leq 0$, we have $x_0\geq \muhat$ and 
\begin{align*}
&\int_{x_0}^{+\infty}\left[\exp\left(V(x;k,\muhat)+a_0\right)-1\right]f(x;\muhat,\shat^2)dx\\
\leq& \int_{x_0}^{+\infty}\left[\exp\left(V(x;k,\muhat)\right)-1\right]f(x;\muhat,\shat^2)dx\\
\leq& \int_{\muhat}^{+\infty}\left[\exp\left(V(x;k,\muhat)\right)-1\right]f(x;\muhat,\shat^2)dx
\end{align*}

Similarly, if $a_0\geq 0$, we have
$$
\int^{x_0}_{-\infty}\left[1-\exp\left(V(x;k,\muhat)+a_0\right)\right]f(x;\muhat,\shat^2)dx\leq \int^{\muhat}_{-\infty}\left[1-\exp\left(V(x;k,\muhat)\right)\right]f(x;\muhat,\shat^2)dx
$$
To summarize, we have
\begin{align*}
d_{TV}(g(x;k,\mu,s^2),f(x;\muhat,\shat^2)) \leq \max&\left( \int_{\muhat}^{+\infty}\left[\exp\left(V(x;k,\muhat)\right)-1\right]f(x;\muhat,\shat^2)dx,\right. \\
&\left.\int^{\muhat}_{-\infty}\left[1-\exp\left(V(x;k,\muhat)\right)\right]f(x;\muhat,\shat^2)dx\right)
\end{align*}

As we have shown in Equation (12) of the main manuscript, $\shat^2 = \left(\frac{2k}{\muhat^2}+C\right)^{-1}$, where $C>0$. This suggests that $\shat\leq \muhat/\sqrt{2k}$. Therefore
\begin{align*}
d_{TV}(g(x;k,\mu,s^2),f(x;\muhat,\shat^2)) \leq \max&\left( \int_{\muhat}^{+\infty}\left[\exp\left(V(x;k,\muhat)\right)-1\right]f(x;\muhat,\muhat/2k)dx,\right. \\
&\left.\int^{\muhat}_{-\infty}\left[1-\exp\left(V(x;k,\muhat)\right)\right]f(x;\muhat,\muhat/2k)dx\right)
\end{align*}

Since $V(x;\mu,s^2)$ and $f(x;\mu,s^2)$ are location-scale families, the above expression can be made free of $\muhat$ and thus $\mu$ and $s^2$:
\begin{equation}
\begin{aligned}
d_{TV}(g(x;k,\mu,s^2),f(x;\muhat,\shat^2)) \leq \max&\left( \int_{1}^{+\infty}\left[\exp\left(V(x;k,1)\right)-1\right]f(x;1,1/2k)dx,\right. \\
&\left.\int^{1}_{-\infty}\left[1-\exp\left(V(x;k,1)\right)\right]f(x;1,1/2k)dx\right)
\end{aligned}
\label{TV.bound.expr}
\end{equation}

This upper bound on the total variation distance decreases as $k$ increases and it goes to $0$ as $k\to\infty$. This suggests the convergence of the approximating normal distribution to the density family $g$ in total variation sense. We also plot this upper bound as a function of $k$ to verify the conclusion. It is shown in the supplemental Figure \ref{TV.bound.plot}.

\begin{figure}[htbp]
\centering
\includegraphics[scale=0.8]{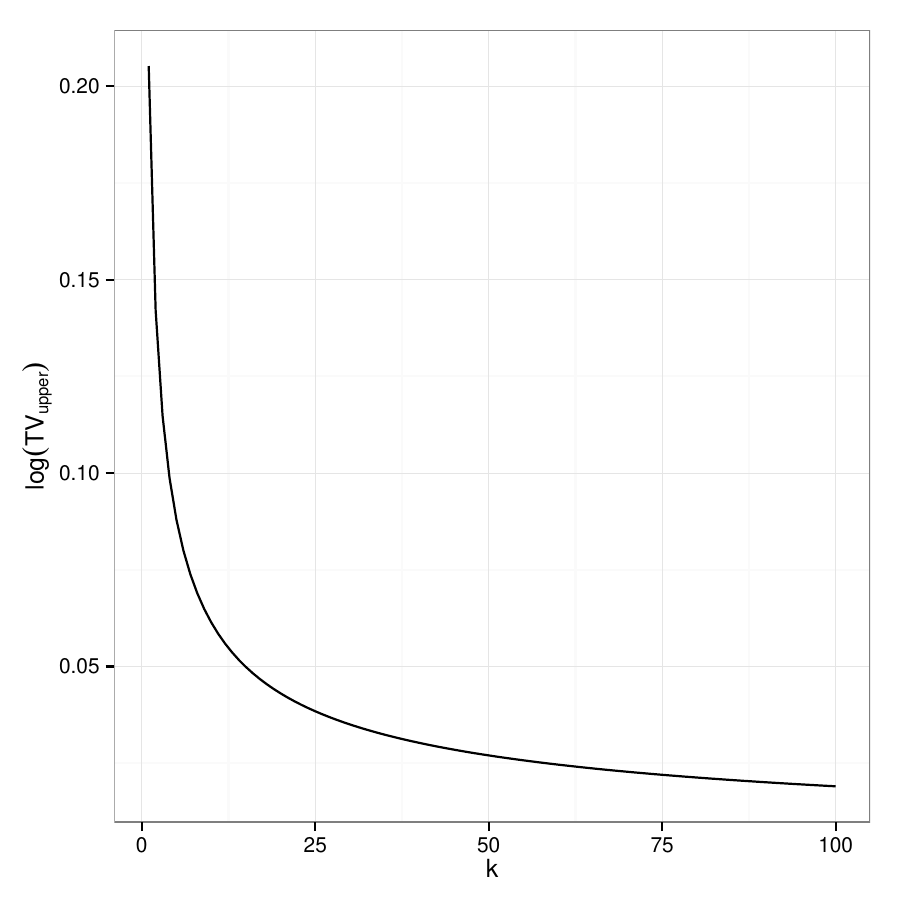}
\caption{Upper bound of the total variation distance of Laplace approximation in (12) to the density in (11) as given in (\ref{TV.bound.expr}) when frequency $k$ increases.}
\label{TV.bound.plot}
\end{figure}

\section{Details of self-consistent estimates in Section 3.1}
First we estimate $\bsigma$ and then we transform the data $n_{i,j}$ into $\sqrt{n_{i,j}/\sigma_i}$. If $n_{i,j}$ is representative and $\bsigma$ is estimated accurately, we have $\sqrt{n_{i,j}/\sigma_i} = c_jQ_{i,j}^{+}$. If the covariance matrix of $\bQ_i$ is $\bSigma$, then the covariance matrix of $(\sqrt{n_{i,j}/\sigma_i},j=1,\ldots,J)$ will be $\bSigmatilde=\bLambda\bSigma\bLambda$ where $\bLambda = diag\{c_1,\ldots,c_J\}$.

It is obvious that $(\sqrt{n_{i,j}/\sigma_i},j=1,\ldots,J)$ is MVN and the correlation matrix will be the same as the induced correlation matrix from $\bSigma$. Methods on identifying the covariance matrix using this truncated dataset are abundant and well-studied. One way to do it is the EM algorithm. This estimated covariance matrix will by no means to be the same as $\bSigma$, but the induced correlation matrix will be very close to the true correlation matrix induced by $\bSigma$. Hence if our interest is on estimating correlation matrix, we can just treat $(\sqrt{n_{i,j}/\sigma_i},j=1,\ldots,J)$ as the truncated version of the true $\bQ_i$ and proceed.

The EM algorithm should then be derived for the following settings. Let $\bQ_i\overset{iid}{\sim} MVN(\bzero,\bSigma)$. Instead of observing $I$ independent $\bQ_i$, we only observe the positive entries in each $\bQ_i$ and know the rest of the entries are negative. Denote the observed data vector as $\bQtilde_i$. We want to estimate $\bSigma$ from the data $\bQtilde_i,i=1,\ldots,I$. A standard EM algorithm can be easily formulated as following:
\begin{enumerate}

\item[E-step] Get the conditional expectation of full data log likelihood, given the observed data. Define two index sets, $\Asc_i=\{j|\Qtilde_{i,j}>0\}$ and $\Bsc_i = \{j|\Qtilde_{i,j}=0\}$. For an arbitrary index set $\Isc$, denote $Q_{\Isc} = (Q_{i,j}|j\in \Isc)$. Denote $\Asc=\{(i,j)|j\in\Asc_i,i=1,\ldots,I\}$ and $\Bsc=\{(i,j)|j\in\Bsc_i,i=1,\ldots,I\}$. The E-step function at $t+1$ iteration is,
$$
L(\Sigma|\Sigma_t) = \mathbb E\left[-\frac{I}{2}\log|\bSigma| - \frac{1}{2}\text{Tr}(\bSigma^{-1}\sum_i\bQ_i\bQ_i')|\bSigma_t,Q_{\Asc}=\Qtilde_{\Asc},Q_{\Bsc}<0\right].
$$

Notice this expectation is not easy to calculate in general. We use instead Monte Carlo method to approximate it. We sample $K$ copies of $\bQ_i$ from the conditional distribution $(\bQ_i|Q_{\Asc_i}=\Qtilde_{\Asc_i},Q_{\Bsc_i}<0)$ where $\bQ_i\sim MVN(\bzero,\bSigma_t)$. The conditional distribution is a truncated multivariate normal distribution and we use the R package \texttt{tmvtnorm} \citep{tmvtnorm} to sample from it. If we denote by $\bQ_i^1,\ldots,\bQ_i^K$ the $K$ samples of $Q_{i}$, $L$ can be approximated as
$$
\Lhat(\Sigma|\Sigma_t) = -\frac{1}{K}\sum_{k=1}^K\left[\text{Tr}(\bSigma^{-1}\sum_i\bQ_i^k(\bQ_i^k)')  \right] -\frac{I}{2}\log|\bSigma|.
$$

\item[M-step] We seek to maximize $\Lhat$ with respect to $\bSigma$. Due to a well-known fact on the maximum likelihood estimate of covariance matrix of multivariate normal, it is straightforward to get
$$
\bSigma_{t+1} = \frac{1}{IK}\sum_{i,k}\bQ_i^k(\bQ_i^k)'.
$$
\end{enumerate}

\begin{figure}[htbp]
\centering
\includegraphics[scale=0.4]{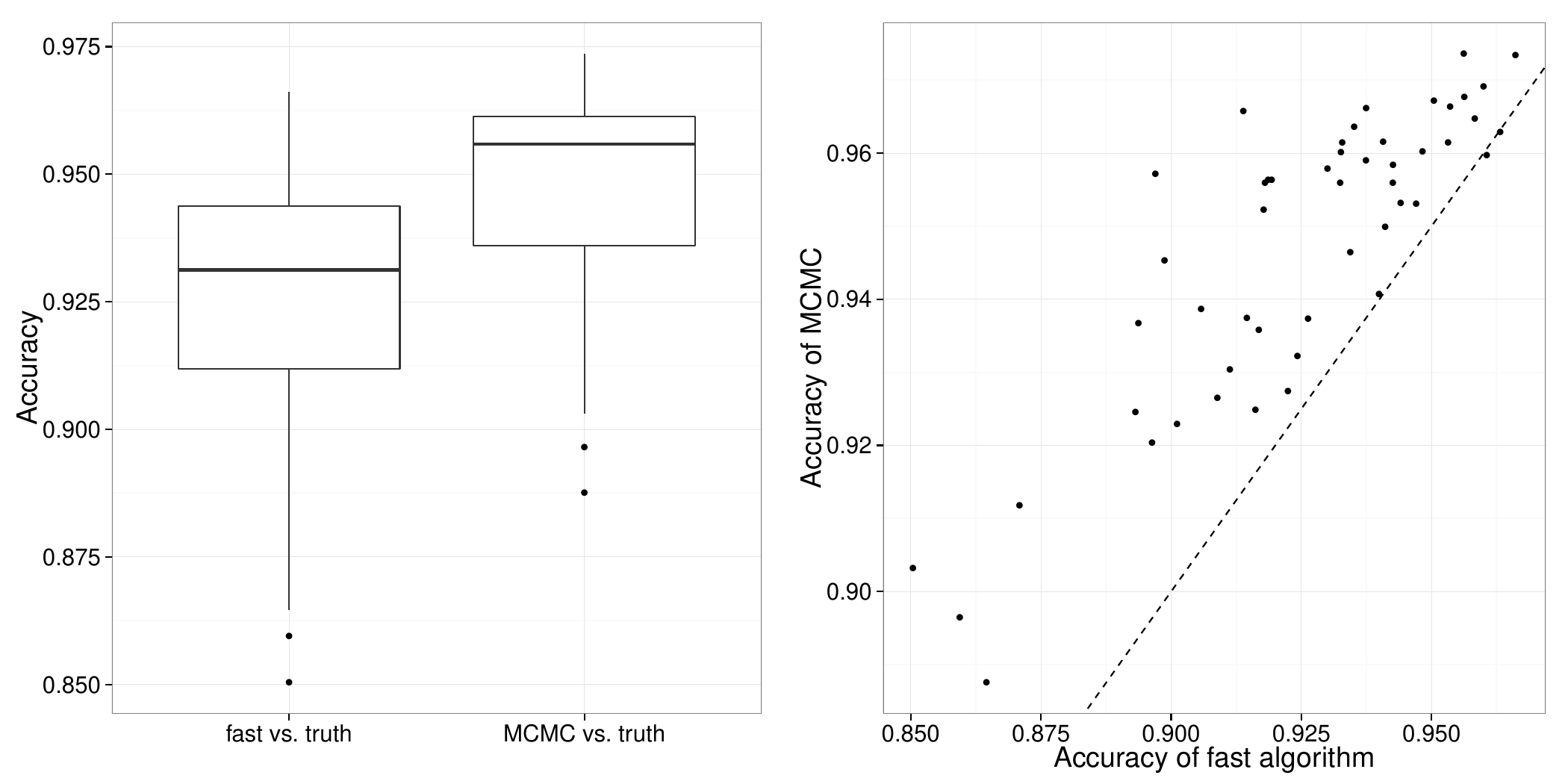}
\caption{(\textbf{Left}) Box-plots compare the distributions of RV-coefficients between estimates from our self-consistent algorithm and between estimates from MCMC simulation and truth. (\textbf{Right}) Scatter plot to show per simulation comparison of RV coefficients for the self-consistent algorithm and MCMC sampling. Dashed line indicates where the two algorithms have identical accuracy.}
\label{fast_check}
\end{figure}

We applied this algorithm to the simulated datasets generated for Figure 3(a) to estimate the normalized Gram matrix $\bS$. A summary of the RV-coefficients between the estimates from the above algorithm and the truth is shown in Figure \ref{fast_check}. We also compared the estimates from this algorithm with those from MCMC simulations in Figure \ref{fast_check}. The estimates of $\bS$ from MCMC simulation are always better than those given by the self-consistent algorithm but both perform very well.

\section{Standard PCoA for ordination of simulated dataset, Global Patterns dataset and Ravel's vaginal microbiome dataset}

\begin{figure}[htbp]
\centering
\includegraphics[scale=0.8]{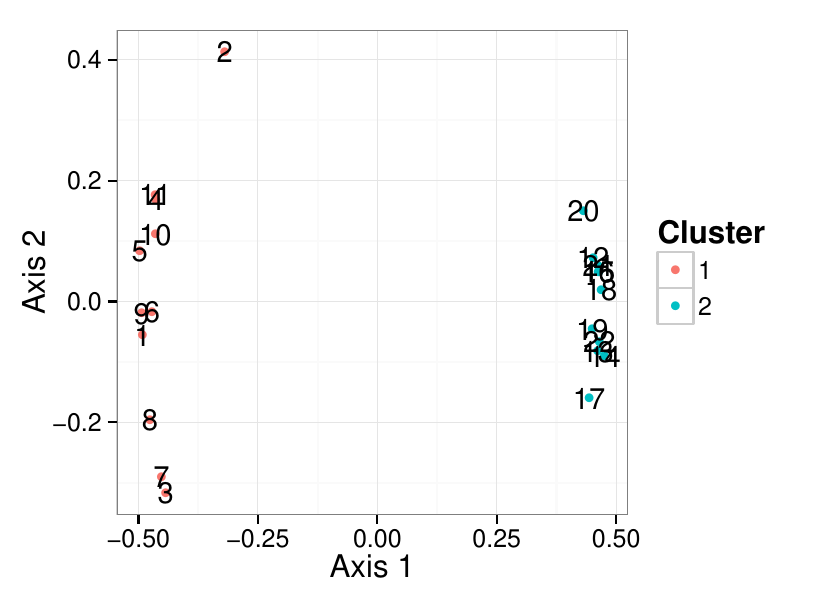}
\caption{PCoA result for the simulated dataset generated for Figure 3(f).}
\label{sim.pcoa}
\end{figure}

\begin{figure}[htbp]
\centering
\includegraphics[scale=0.65]{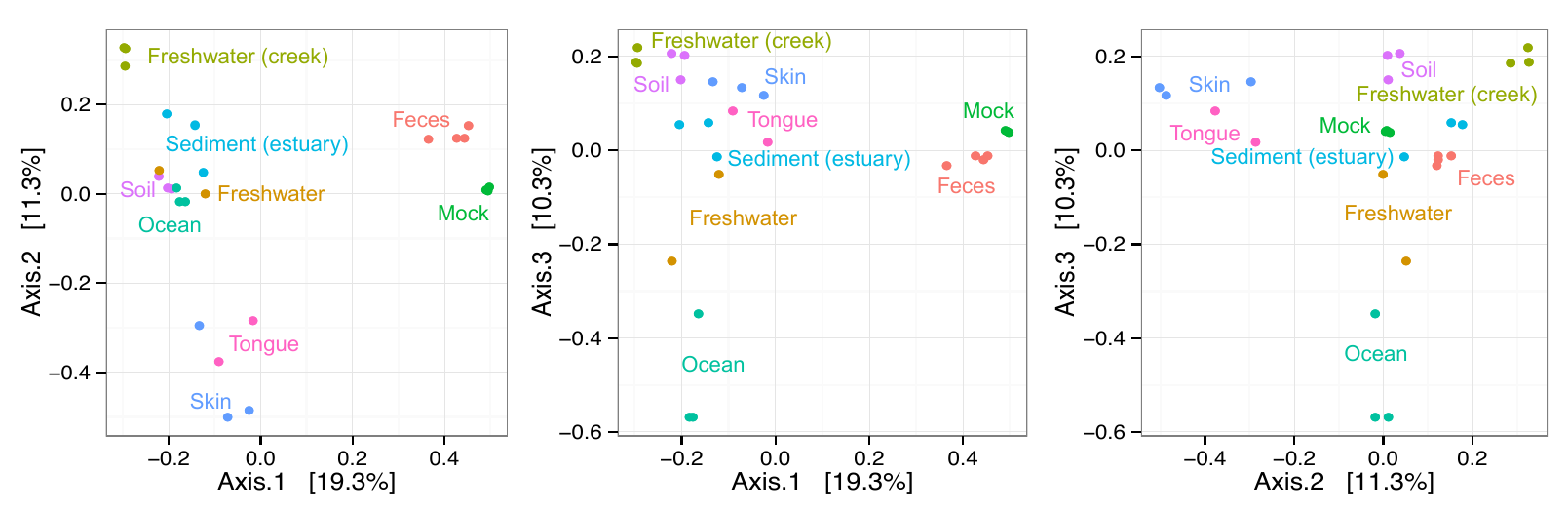}
\caption{PCoA results for the Global Patterns dataset. We show the three two-dimensional representations of the ordination given by the first three principal coordinates.}
\label{sim.gp}
\end{figure}

\begin{figure}[htbp]
\centering
\includegraphics[scale=0.28]{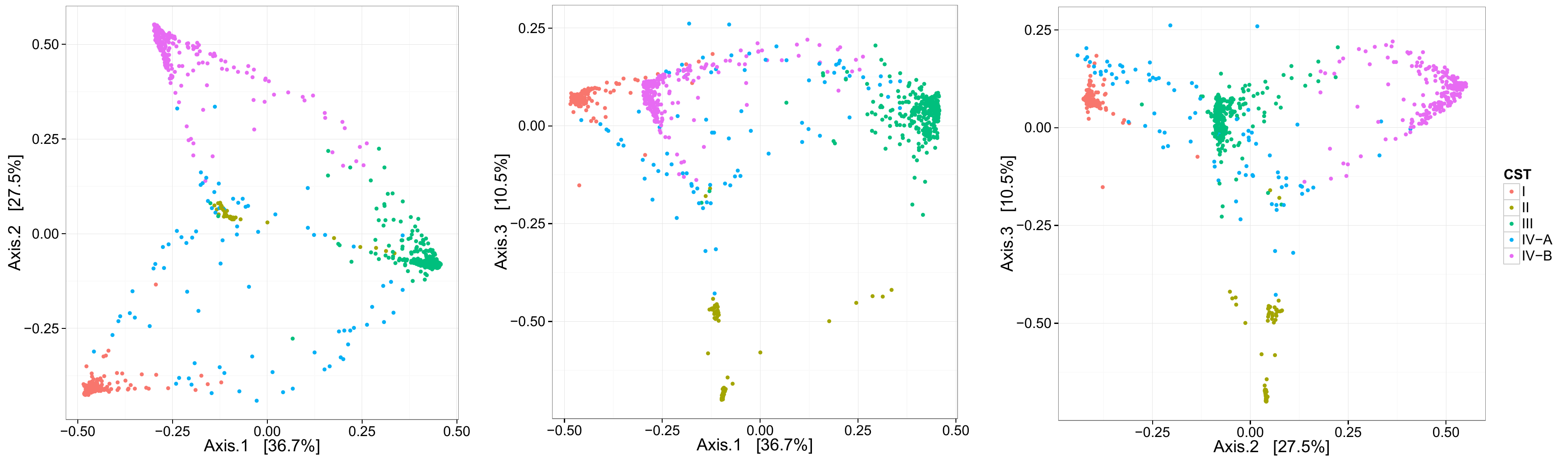}
\caption{PCoA results for Ravel's vaginal microbiome dataset. We show the three two-dimensional representations of the ordination given by the first three principal coordinates.}
\label{sim.ravel}
\end{figure}

In this section, we include three sets of ordination figures generated using the standard PCoA method in microbiome studies. We first calculate the dissimilarity matrix of biological samples by applying Bray-Curtis dissimilarity metric on the empirical microbial distributions. We then perform classic Multi-dimensional Scaling (MDS) to ordinate biological samples based on the dissimilarity matrix. In Figure \ref{sim.pcoa}, we show the PCoA result for the simulated dataset generated for Figure 3(f). In Figure \ref{sim.gp} and \ref{sim.ravel}, we illustrate the PCoA results for the Global Patterns dataset and Ravel's vaginal microbiome dataset respectively. To be consistent with the main results, we show the ordination results based on the first three principal coordinates for the Global Patterns dataset and Ravel's vaginal microbiome dataset.

\section{Benchmarking the MCMC sampler}
In this section, we focus on evaluating the computational performance of our MCMC sampler. We first consider the computational time of the sampler under different scenarios. We then illustrated a convergence diagnosis to check whether  the sampler has reached mixing in the setting of our simulation study in the main manuscript. In addition, we created two larger datasets to verify the number of iterations needed to reach mixing will not be compromised if the underlying latent structure remains low dimensional.

\subsection{Computation time of the MCMC sampler}
In Table \ref{comp.time} we listed the elapsed time in seconds for the MCMC sampler to finish $1,000$ iterations under different scenarios. All the scenarios are run with a single thread on a MacBook Pro with 2.7GHz Intel Core i5 and 8 GB 1867 MHz DDR3 RAM. In particular, we evaluated the effect of the number of biological samples ($J$), the number of species ($I$), the dimension of the latent factors ($m$), and the total counts per biological sample ($n^j$).

\begin{table}[htbp]
\footnotesize
  \centering
  \caption{Computation time (in seconds) of 1,000 iterations for the MCMC sampler}
    \begin{tabular}{ccccccccccc}
    \toprule
    \multicolumn{2}{c}{\multirow{2}[4]{*}{}} & \multicolumn{3}{c}{$I=68$} & \multicolumn{3}{c}{$I=500$} & \multicolumn{3}{c}{$I=1000$} \\
\cmidrule{3-11}    \multicolumn{2}{c}{} & $m=5$ & $m=10$ & $m=20$ & $m=5$ & $m=10$ & $m=20$ & $m=5$ & $m=10$ & $m=20$ \\
    \midrule
    \multirow{3}[2]{*}{$J=22$} & $n^j=10^3$ & 2.3   & 2.8   & 2.4   & 5.7   & 5.8   & 7.0   & 11.4  & 10.4  & 12.6 \\
          & $n^j=10^4$ & 1.3   & 1.6   & 1.9   & 5.7   & 5.5   & 6.4   & 8.7   & 8.8   & 11.3 \\
          & $n^j=10^5$ & 1.1   & 1.4   & 1.5   & 4.7   & 3.9   & 6.3   & 7.2   & 8.2   & 11.5 \\
    \midrule
    \multirow{3}[2]{*}{$J=100$} & $n^j=10^3$ & 3.6   & 3.7   & 5.5   & 11.5  & 14.6  & 17.1  & 21.8  & 21.0  & 30.2 \\
          & $n^j=10^4$ & 3.3   & 3.7   & 5.4   & 11.5  & 12.1  & 20.4  & 18.1  & 21.1  & 29.5 \\
          & $n^j=10^5$ & 3.4   & 4.0   & 5.5   & 12.3  & 18.9  & 17.8  & 19.2  & 21.5  & 31.1 \\
    \midrule
    \multirow{3}[2]{*}{$J=1000$} & $n^j=10^3$ & 31.4  & 34.3  & 49.6  & 121.2 & 118.4 & 152.1 & 152.1 & 173.8 & 251.0 \\
          & $n^j=10^4$ & 28.2  & 33.4  & 53.1  & 96.3  & 144.3 & 159.7 & 143.7 & 164.8 & 254.2 \\
          & $n^j=10^5$ & 40.1  & 38.2  & 52.2  & 129.1 & 111.5 & 138.2 & 163.2 & 171.7 & 246.0 \\
    \bottomrule
    \end{tabular}%
  \label{comp.time}%
\end{table}%
Increasing the total number of reads per biological sample ($n^j$) does not affect the computation time. On the other hand, there is a weak effect associated with the dimension of the latent factors ($m$). In general, the computation time tends to increase with $m$. The number of species ($I$) and the number of biological samples ($J$) affect the speed of computation significantly. These results illustrate that the MCMC sampler can finish $50,000$ iterations for a dataset with 100 samples and 1000 species in less than 20 minutes.

The table illustrates that it is possible to apply our model to microbiome datasets with comparable numbers of biological samples. It is rare to have datasets with more than a thousand confidently assigned OTUs \citep{DADA2}.

\subsection{Convergence diagnosis of the MCMC sampler}
\label{chap.conv}
\begin{figure}[htbp]
\centering
\includegraphics[scale=0.4]{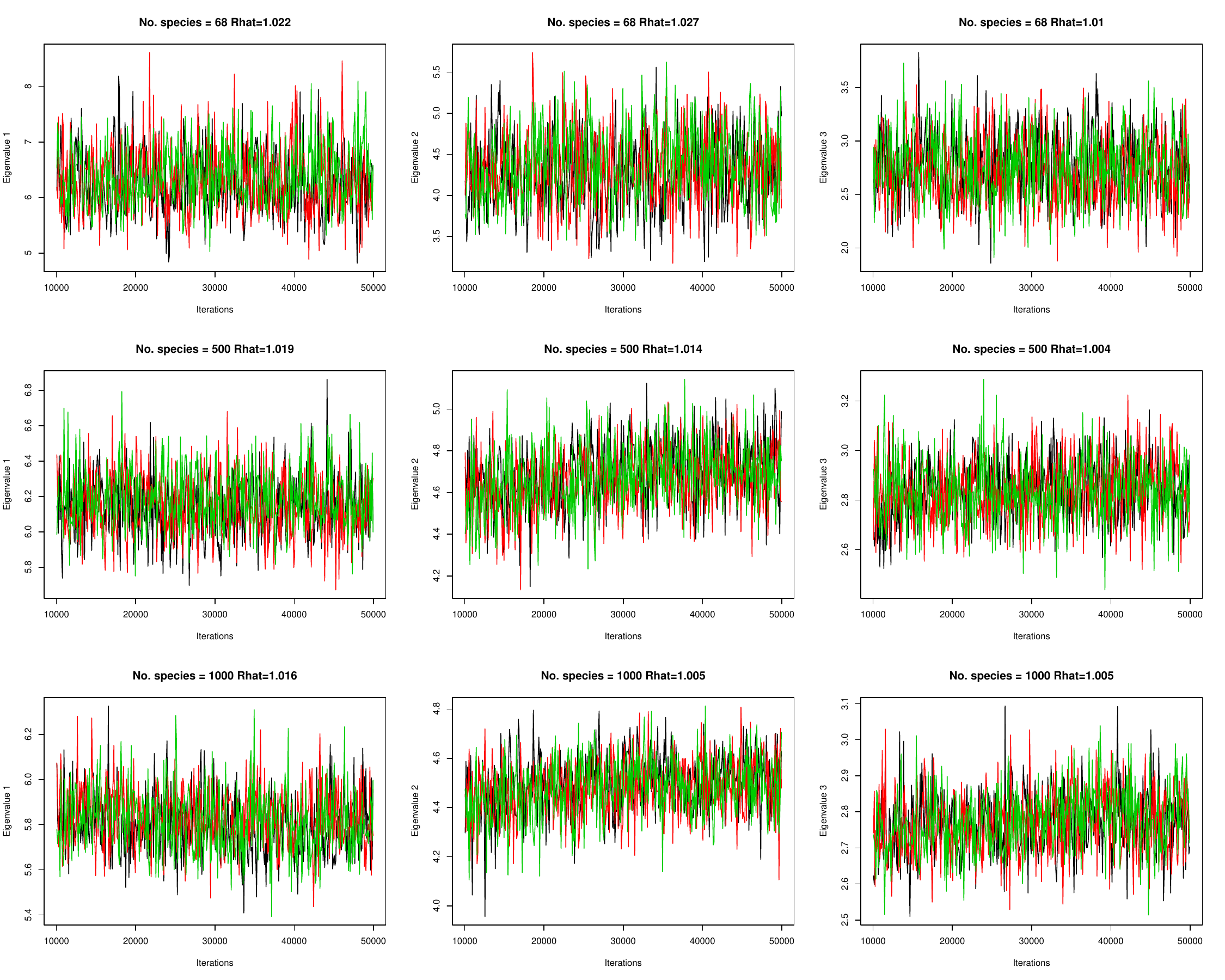}
\caption{Traceplots for the posterior samples of the first three eigenvalues of $\bS$. Each row corresponds to a different $I$ and each column to a different eigenvalue. The $\Rhat$ statistics are shown in the title of each figure.}
\label{mixing.figs}
\end{figure}

We evaluate the convergence of the MCMC sampler in the setting of Section 5 (simulation study). The number of biological samples is fixed at $J = 22$. We ran three parallel chains for three scenarios $I = 68$, $I = 500$ and $I = 1,000$. For each different $I$, we obtain the posterior samples of the first three eigenvalues of the normalized Gram matrix $\bS$ in all three chains and use $\Rhat$  statistics \citep{gelman.check} to check if the chains reached mixing. We chose to visualize the eigenvalues of $\bS$ since in our model $\bS$ is identifiable. The results are shown in Figure \ref{mixing.figs}.

The $\Rhat$ statistics are all close to one supporting good MCMC mixing after 20,000 iterations, so our choice of 50,000 total iterations seems reasonable for providing posterior inference.

\end{appendices}

\bibliographystyle{chicago}

\bibliography{reference}

\end{document}